\documentclass[aps,preprint,groupedaddress,showpacs]{revtex4}
\usepackage{graphicx}

\def\kappav{{\mbox{\boldmath{$\kappa$}}}}
\def\rhov{{\mbox{\boldmath{$\rho$}}}}
\begin{document}
% You should use BibTeX and apsrev.bst for references
\bibliographystyle{apsrev}

%\preprint{}
%
%%
%%
% FOR TWO COLUMN  ACTIVATE THE LINE BELOW 
%\twocolumn[\hsize\textwidth\columnwidth\hsize\csname @twocolumnfalse\endcsname
\title{QUANTUM DYNAMICS OF A HYDROGEN MOLECULE CONFINED IN A CYLINDRICAL POTENTIAL}
\author{Taner Yildirim}
\affiliation{National Institute of Standards and Technology,
Gaithersburg, MD 20899}
\author{A. B. Harris}
\affiliation{Department of Physics and Astronomy, University of Pennsylvania,
Philadelphia, PA 19104}
\date{\today}
\begin{abstract}
We study the coupled rotation-vibration  levels of a hydrogen molecule in 
a confining potential with cylindrical symmetry.
% when the rotational quantum number $J$ is a good quantum number. 
We include the coupling between rotations and translations and show how this
interaction is essential to obtain the correct degeneracies of the energy level
scheme. We applied our formalism to study the dynamics of H$_{2}$ molecules
inside a "smooth" carbon nanotube as a function of tube radius.
The results are obtained both by numerical solution of the 
($2J+1$)-component radial Schrodinger equation and by developing
an effective Hamiltonian to describe the splitting of a manifold
of states of fixed angular momentum $J$ and number of phonons, $N$.
For nanotube radius smaller than $\approx 3.5$ \AA, the confining
potential has a parabolic shape and the results can be understood
in terms of a simple toy model. For larger radius, the potential
has the "Mexican hat" shape and therefore the H$_{2}$ molecule is
off-centered, yielding radial and tangential translational dynamics 
in addition to rotational dynamics of H$_{2}$ molecule which we also
describe by a simple model.  Finally, we make several predictions for the
the neutron scattering observation of various transitions between
these levels.
\end{abstract}
%%
%% 82.80.Gk Analytical methods involving vibrational spectroscopy
%% 34.50.Ez Rotational and vibrational energy transfer
%% 36.20.Ng Vibrational and rotational structure, infrared and Raman spectra
%%63.22.+m Phonons or vibrational states in low-dimensional structures and nanoscale materials
%%25.40.Fq Inelastic neutron scattering
%%78.70.Nx Neutron inelastic scattering
%%71.20.Tx Fullerenes and related materials; intercalation compounds .
\pacs{78.70.Nx,34.50.Ez,82.80.Gk,71.20.Tx,36.20.Ng,63.22.+m}
\maketitle

\section{INTRODUCTION}

The study of quantum dynamics of hydrogen molecules in confined
geometries has recently developed into an active field both
experimentally and 
theoretically\cite{fitzgerald,ramanprl,narehood,kostov,TYABH,gs1,gs2,gs3,price,brown,NIST,novaco}
due to potential use as catalysts, molecular sieves, and storage media.
In the case of fullerenes and nanotubes, such trapping may
yield new exotic quantum systems due to zero and one dimensionality
of the absorption sites, respectively. Thus, understanding the
structural and dynamical aspects of trapping in confining 
geometries is of both fundamental and practical importance.

The theory of molecular rotation in solids has a long history
dating back to the early work of Pauling\cite{pauling}, 
Devonshire\cite{devonshire}, and Cundy\cite{cundy}. 
They introduced the concept of the crystal field potential,
$V(\Omega)$, where  $\Omega$ specifies the orientation of the
molecule, to solve for the energy levels of the hindered rigid-rotor.
This traditional approach assumes that the center of mass (CM) 
of the trapped molecules are fixed and therefore does not take into
account the  rotation-vibration (RV) coupling. However recent
studies\cite{TYABH,NIST} have indicated that vibrational levels of 
H$_{2}$ trapped
in the octahedral sites of C$_{60}$, for example, are
significantly perturbed by RV coupling and in a previous paper\cite{TYABH} (I)
we have shown that this coupling
has to be included in a symmetry analysis of the energy level degeneracies.
Interestingly, to date there is a little done to treat CM dynamics and 
RV coupling. Most of the studies are based on the approximation 
where an effective orientational crystal field potential is
obtained after the potential is averaged over the zero-point
translational motions of the H$_{2}$ molecule\cite{NIST,novaco}.

Recently (in I) we have presented a detailed theory of coupled
RV dynamics of H$_{2}$ molecule trapped in
a zero dimensional cavity with various symmetries. Here we
present a similar study to analyze the combined rotational 
and translation states of hydrogen molecules  
confined in a one dimensional potential. This problem is closely
related to the experimental situation where hydrogen molecules
are absorbed  into carbon nanotube ropes. Figure~1 shows schematically
various types of absorption sites for H$_{2}$ molecule. Several 
neutron and Raman scattering experiments have been carried out 
to characterize the binding energies and rotational barriers 
for H$_{2}$ at these sites with conflicting 
results\cite{fitzgerald,ramanprl,narehood,kostov,price,brown,NIST}. 
One of the motivations
of the present work is to provide a detailed description of the
RV dynamics of H$_{2}$ molecules at these
different absorption sites and discuss the consequences for
inelastic neutron scattering experiments. In the present paper,
we focus our attention on the general formalism and discuss only
the case where a single H$_{2}$ is confined inside a single nanotube.
Extension of this work to the interstitial and external sites and
to cases where H$_2$ molecules interact with one another will
be presented elsewhere.

Briefly this paper is organized as follows. In the next section,
we discuss the potential model for hydrogen and nanotube interactions
and validate several approximations, such as assuming a smooth tube,
used in our formalism. In Sec. III we present our formalism to treat
the coupled rotational and translational motion of H$_{2}$ molecule
confined in a smooth nanotube. We show that the problem can
be mapped into a ($2J+1$)-component radial Schrodinger equation
which can be solved numerically. In Sec. IV we discuss the
dynamics of a hydrogen molecule when the confining potential has
parabolic shape (which occurs for a small-radius nanotube). 
We interpret the exact numerical results in terms of a simple
analytical toy model. In Sec. V we discuss the case where
the confining potential has a Mexican-hat shape. For this case
(which occurs for large radius nanotubes such as (10,10)) the
equilibrium position of the CM of the H$_{2}$ molecule is 
off-center and it performs radial and tangential translational
oscillation in combination  with its rotational dynamics. 
In this section, we also present several perturbation results
which help to interpret the exact numerical results. In Sec.VI
we discuss the experimental observation of various transitions
via inelastic neutron scattering measurements. Our conclusions
are summarized in Sec.VII.

\section{POTENTIAL MODEL}

We model the intermolecular potential for H$_{2}$ trapped in a carbon
nanotube as a sum of atom-atom potentials

\begin{equation}
V({\bf r} , \Omega) = \sum_{i,H}\sum_{j,C} \biggl(
B \exp(-C r_{ij}) - A/r_{ij}^{6} \biggr)
\; .
\end{equation}
The dependence of the potential on the position (${\bf r}$) and 
orientation ($\Omega$) of H$_{2}$ molecule is through the inter-atomic
distances $r_{ij}$.
All the results reported in this paper
are obtained from the same WS77 potential,\cite{ATOM}
$ -A/r^{6}+B \exp(-Cr)$, 
(where $A=5.94\ {\rm eV} \AA^{6}, B= 678.2$  eV, and $C=3.67 \AA^{-1}$),
that we  used in I\cite{TYABH}. Compared to other commonly used
potentials, the WS77 potential gave the best fit to the energy
spectrum of H$_{2}$ in solid C$_{50}$.

For simplicity we will restrict this formulation to the idealized 
case when the hydrogen
molecule is confined by a so-called ``smooth" nanotube.  By this we mean that the
potential produced by the nanotube has cylindrical symmetry and is invariant with
respect to translations along its axis of symmetry.  In some of our numerical work
we will study ``real'' nanotubes which 
do not posses the high symmetry of ``smooth'' nanotubes.

It is instructive to look at various potentials for an orientationally 
averaged hydrogen molecule (i.e. para hydrogen with $J=0$) when H$_{2}$
is inside and outside a single nanotube. 
Figure~2 indicates two different types of confining potential depending on
the nanotube radius. Figure ~2a shows that for small nanotubes such as (9,0),
the potential minimum occurs at the center of the tube and therefore
the potential has a parabolic shape. However for larger nanotubes
such as (10,10) nanotube, the minimum is off-centered and therefore the
potential has a Mexican-hat shape. Because of this off-centering
the dynamics of the H$_{2}$ molecule is a quite interesting and rich one as
we discuss in detail below. The right panels in Fig.~2 shows the 
potential when the H$_{2}$ molecule is outside the nanotubes.
The outside binding energy does not depend on the tube radius strongly
and is about 30 meV. The horizontal lines indicates the 
radial phonon energy levels, indicating that at least a few bound states
can occur even for a hydrogen molecule outside a single nanotube.
The solid and dashed lines in Fig.~2 shows the results with and without
the  smooth tube approximation, respectively. Since these two
curves are very close to one another, the smooth tube approximation will
not cause significant error in our theory. 

In order to develop some intuition about the orientational potential for 
a hydrogen molecule in a nanotube, in Fig.~3a we show the radial potential 
for three different orientations of H$_{2}$ molecule inside a 
(10,10) nanotube. We note that the radius of the (10,10) tube is 
large enough that the parallel to the axis (p) and tangential (t) orientations
(as depicted in the inset to Fig.~3a) give almost the same energy.
However, the radial (r) orientation of a H$_{2}$ molecule has 
a minimum energy which is about 8 meV higher in energy than that of
the other two orientations. We also
point out that the position of the CM of the H$_2$ molecule for the
minimum potential energy changes about 0.2 $\; \AA$
depending on the orientation of the
H$_{2}$ molecule. This is a clear indication that the orientational
and vibrational motion of hydrogen molecule are significantly
coupled.

Figure~3b shows the minimum potential energies, $E_{r}, E_{p}$, and 
$E_{t}$ for respective radial, parallel, and tangential
orientations of an H$_{2}$
molecule inside various nanotubes. It is clear that for nanotube
radius around 3$\;$\AA, the orientational dependence of the
potential is of the order of 30 meV and therefore is comparable
to the energy separation of 80 meV or more between energy levels
corresponding to different rotational quantum number $J$'s.
Accordingly, for tubes whose radius is less than that of
a (9,0) tube, $J$ may not be considered to be a good quantum
number. For large nanotubes,
such as (10,10), the orientational potential is of the order
of 8 meV and does not change much with larger radius tubes
(i.e. close to the graphite limit). In the present paper,
we present our formalism using (9,0) and (10,10) nanotubes
which represent the two potential regimes; namely the
parabolic and Mexican-hat potentials, but for both tubes
$J$ is considered to be a good quantum number.

\section{FORMULATION}

The hydrogen molecule is unique in that its moment of inertia
is small enough that the rotational kinetic energy
often dominates the orientational potential in which the molecule is placed.
Under these circumstances the rotational quantum number $J$ is nearly a good
quantum number and the effect of the orientational potential is to reduce the
degeneracy of the $2J+1$ sub-states of a given $J$.  
(The generalization of the formulation we present below to the 
case when $J$ is not a
good quantum number will be presented elsewhere\cite{TYABH3}).
In the present case any eigenfunction describing the
orientational and translation state of the molecule can be written in the form
\begin{eqnarray}
\Psi({\bf r} , \Omega) &=& \psi(r,\phi_r;\Omega)e^{ikz} 
= \sum_{M=-J}^J \Phi_M^{J,k} (r , \phi_r) Y_J^M(\Omega) e^{i k z} \ ,
\label{PSIEQ} \end{eqnarray}
where $r$, $z$, and $\phi_r$ are the cylindrical coordinates of the
center-of-mass
of the hydrogen molecule, $\Omega$ denotes its molecular orientation specified
by angles $\theta$ and $\phi$, and $Y_J^M(\Omega)$ is a spherical harmonic.
We will refer to $\psi(r,\phi_r;\Omega)$ as the cylindrical RV 
wavefunction.  For economy of notation we henceforth omit the
superscripts $J$ and $k$.  Because $\Phi_M$ is allowed to depend
arbitrarily on $r$ and $\phi_r$, this wave function takes into
account the most general interaction between rotations and
translations subject to the constraint that $J$ is a good
quantum number.  In this notation, the Hamiltonian for a single
hydrogen molecule (of mass $m$) with $J$ and $k$ fixed is written as
\begin{eqnarray}
{\cal H} &=& - {\hbar^2 \over 2m} \nabla^2 + B J(J+1) + V({\bf r}, \Omega) \ ,
\end{eqnarray}
where $m$ is the mass of an H$_2$ molecule and
$V({\bf r},\Omega)$ is the orientational potential which, as indicated, 
also depends on position.   For a smooth nanotube we may write the orientational
potential energy as
\begin{eqnarray}
V({\bf r},\Omega) = V_0 (r)
+ \sum_{L,M} V_L^M(r , \phi_r) Y_L^M( \theta , \phi) \ ,
\end{eqnarray}
where the sum is over $L>0$.  Because the
hydrogen molecule is centrosymmetric, only terms with $L$ even appear in the 
potential.  Also, because a smooth nanotube has a mirror plane perpendicular to
the axis of symmetry, only terms with even $M$ appear.
Furthermore,  because a global rotation of the molecule
(i. e. incrementing both $\phi$ and $\phi_r$ by
the same amount) is a symmetry of the system, we may write
\begin{eqnarray}
V({\bf r},\Omega) = V_0 (r) + \sum_{L,M} v_L^M(r) Y_L^M( \theta ,0)
e^{i M(\phi-\phi_r)} \ ,
\label{VIBROT} \end{eqnarray}
where $v_L^M$ is a function only of $r$ and $v_L^{-M}={v_L^M}^*$. 
In addition, $v_L^M(r=0)$ vanishes for $M \not= 0$.
There is also a mirror plane containing the long axis of the tube which
implies that the potential should be an even function of $(\phi-\phi_r)$.
This implies that $v_L^M(r)$ is a real-valued function.  This function
may be evaluated by integrating the potential at a
fixed center-of-mass position over all orientations:
\begin{eqnarray}
v_L^M(r) = e^{i M \phi_r} \int d \Omega Y_L^M(\theta, \phi)^*
V({\bf r},\Omega) \ .
\end{eqnarray}
Contrary to appearance, $v_L^M(r)$ does not depend on
$\phi_r$ because $V({\bf r},\Omega)$ is a function of
$(\phi-\phi_r)$.

The Schrodinger equation for $\psi(r,\phi_r;\Omega)$ is
\begin{eqnarray}
&& \Biggl[ - {\hbar^2 \over 2m} \Biggl( {\partial^2 \over \partial r^2}
+ {1 \over r} {\partial \over \partial r}
+ {1 \over r^2} {\partial^2 \over \partial \phi_r^2} \Biggr) 
+ V_0 (r) + \sum_{LM} v_L^M (r ) e^{-iM \phi_r} Y_L^M ( \Omega )
\Biggr] \psi^{(\alpha)} (r, \phi_r ; \Omega )
\nonumber \\ && \ \
= \hat E^{(\alpha)}  \psi^{(\alpha)} (r, \phi_r ; \Omega) \ ,
\label{SCHEQ} \end{eqnarray}
where $\hat E^{(\alpha)} = E^{(\alpha)} - BJ(J+1) - \hbar^2 k^2/(2m)$. 
For given values of $J$ and $k$, this equation generates a spectrum
of eigenvectors $\psi^{(\alpha)}(r, \phi_r; \Omega)$ with associated
eigenvalues $\hat E^{(\alpha)}$, for $\alpha=0, 1, \dots$.
Substituting Eq. (\ref{PSIEQ}) into
the Schrodinger equation we rewrite it in the form
\begin{eqnarray}
&& \Biggl\{ - {\hbar^2 \over 2m} \Biggl( {\partial^2 \over \partial r^2}
+ {1 \over r} {\partial \over \partial r}
+ {1 \over r^2} {\partial^2 \over \partial \phi_r^2} \Biggr) 
+ V_0 (r) \Biggr\} \Phi_M^{(\alpha)} (r , \phi_r) \nonumber \\ && \ 
+ \sum_{L, M'} \Biggl( \int d \Omega Y_J^M(\Omega)^* Y_L^{M'} (\Omega)
Y_J^{M-M'} (\Omega) d \Omega \Biggr) v_L^{M'}(r)
\Phi_{M-M'}^{(\alpha)} (r , \phi_r) e^{-iM' \phi_r}\nonumber \\ &&
\ = \hat E^{(\alpha)} \Phi_M^{(\alpha)} (r, \phi_r) \ , \;\;\;
M=-J, -J+1 \dots J-1, J \ .
\end{eqnarray}
We see that we may write a solution to this set of equations in the form
\begin{eqnarray}
\Phi_M^{(\alpha)} (r, \phi_r) &=& f_{M,P}^{(\alpha )} (r)
e^{-i(P+M)\phi_r} \ ,
\end{eqnarray}
where $P$ is a quantum number whose significance we will discuss shortly.  
Thus we have
\begin{eqnarray}
&& \Biggl\{ - {\hbar^2 \over 2m} \Biggl( {\partial^2 \over \partial r^2}
+ {1 \over r} {\partial \over \partial r} - {1 \over r^2} (P+M)^2 \Biggr) 
+ V_0 (r) \Biggr\} f_{M,P}^{(\alpha)} (r ) \nonumber \\ && \ 
+ \sum_{L, M'} v_L^{M'}(r) f_{M-M',P}^{(\alpha)}(r)
\Biggl( \int d \Omega Y_J^M(\Omega)^* Y_L^{M'} (\Omega)
Y_J^{M-M'} (\Omega) d \Omega \Biggr) = E f_{M,P}^{(\alpha)} (r ) \ .
\label{EQN} \end{eqnarray}
For each $P$ index we have a Schrodinger equation for the $(2J+1)$-component
wave function which is of the form
$[f_{-J,P}^{(\alpha)}(r), f_{-J+1,P}^{(\alpha)}(r), \dots
f_{J,P}^{(\alpha)}(r)]$.  Then the cylindrical RV wavefunction is
\begin{eqnarray}
\psi_P^{(\alpha)} (r , \phi_r, \Omega) &=& e^{-iP \phi_r}
\sum_M f_{M,P}^{(\alpha)} (r) Y_J^M(\Omega) e^{-iM\phi_r} =
e^{-iP \phi_r} \sum_M f_{M,P}^{(\alpha)} (r) Y_J^M(\theta)
e^{iM(\phi-\phi_r)} \ .
\label{PEQ} \end{eqnarray}
It is important to keep in mind that $v_L^M(r)$ vanishes for odd $M$.
As a consequence, in the $(2J+1)$-component wavefunction there is no mixing
between even and odd values of $M$.  For $J=1$ wavefunctions one will have
``even'' wavefunctions in which the sum over $M$ in Eq. (\ref{PEQ}) reduces
to the single term for $M=0$ and ``odd'' wavefunctions in which the sum over
$M$ in Eq. (\ref{PEQ}) includes only $M= \pm 1$.

The quantum number $P$ indicates that this wavefunction transforms like 
$e^{-iP \phi}$ when the position and orientation of the molecule are
simultaneously rotated about the axis of symmetry.  Under this global
rotation the quantity
\begin{eqnarray}
f_{M,P}^{(\alpha)}(r) e^{-iM\phi_r} Y_J^M(\Omega)
\end{eqnarray}
is invariant, so the total wavefunction transforms as stated.  When
$v_L^M(r)$ is independent of $r$, then, since $v_L^M(r=0)$ must vanish,
we have that $v_L^M(r)=0$ for $M \not= 0$ and, in Eq (\ref{EQN}) there
is no coupling between
$f^{(\alpha)}_M$'s for different values of $M$.  Thus, in this case
there is no dynamical interaction between the orientational coordinate
and the center-of-mass coordinate and
the wave function can be chosen so that $f_{M,P}$ is only nonzero
for a single value of $M$. Then one has the usual separation
of variables so that the orientational wave function is proportional
to $e^{iM\phi}$ and the translation wave function is proportional to
$e^{-i(M+P)\phi_r}$.
Here we have accomplished a similar separation of coordinates
when $v_{L}^{M}(r)$ is allowed to depend on $r$.
Now the result is not a scalar radial equation,
but rather a radial equation for a $(2J+1)$-component wave function.
That is the result embodied in Eq. (\ref{EQN}), where we have one such
$(2J+1)$-component radial Schrodinger equation for each value of $P$.

\section{QUASI-HARMONIC POTENTIAL}

In this section we discuss the case exemplified by a H$_2$ molecule inside
a $(9,0)$ nanotube,
for which the minimum of the potential $V_0(r)$ occurs for $r=0$, in
which case we will introduce a toy model with the isotropic
harmonic potential $V_0(r) = \frac{1}{2} k r^2$. 

\subsection{No Interactions between Rotations and Translations}

\subsubsection{Harmonic Potential}

Here we discuss the eigenvalues and eigenfunctions of the two-dimensional
isotropic harmonic oscillator, to emphasize the relation between the above
formulation in terms of cylindrical coordinates and that in terms of
Cartesian coordinates.

For an isotropic and harmonic potential we expect the eigenvalues to be
\begin{eqnarray}
E_n &=& (n+1) \hbar \omega = (n+1) \sqrt{ {k/m} } \ .
\end{eqnarray}
Note that the $n$th level (with energy $n\hbar \omega$)
is $n$-fold degenerate, because in Cartesian notation, if, say
$n=4$, we have wave functions $(3,0)$, $(2,1)$, $(1,2)$, and $(0,3)$,
where $(n,m)$ is a wave function
with $n$ excitations in the $x$ coordinate and $m$ excitations in the
$y$ coordinate.   This degeneracy reflects the $U_2$ symmetry corresponding
to the invariance of the Hamiltonian with respect to a transformation of
the form
\begin{eqnarray}
\left( \begin{array} {c} (a_x^\dagger)' \\ (a_y^\dagger)' \\ \end{array} \right)
&=& \Biggl[ {\bf U} \Biggr] 
\left( \begin{array} {c} (a_x^\dagger) \\ (a_y^\dagger) \\ \end{array} \right) \ ,
\end{eqnarray}
where $a_x^\dagger$ and $a_y^\dagger$ create phonons in the $x$ and
$y$-coordinates, respectively and ${\bf U}$ is a two-dimensional unitary matrix.
This transformation is essentially the same as a four
dimensional rotational symmetry in the space of
the momenta $p_x$, $p_y$, and coordinates $x$ and $y$.
Since the kinetic energy is quadratic in the momenta, 
spherical symmetry in this space only holds if the
potential is harmonic.

In cylindrical coordinates the eigenfunctions can be written as
\begin{eqnarray}
\psi^\alpha_\mu(r) e^{i \mu \phi} \ ,
\end{eqnarray}
where $\psi^\alpha_\mu(r)$ satisfies the radial equation,
\begin{eqnarray}
- {\hbar^2 \over 2m} \Biggl[ {d^2 \psi^\alpha_\mu(r) \over dr^2}
+ {1 \over r} {d \psi^\alpha_\mu(r) \over dr}
- {\mu^2 \over r^2} \Biggr] + \frac{1}{2} k r^2 \psi^\alpha_\mu(r)
= E^\alpha_\mu \psi^\alpha_\mu(r) \ .
\end{eqnarray}
Here the family of solutions for a given value of $\mu$ are labeled $\alpha=0,1,2, \dots$
in order of increasing energy and for the isotropic and harmonic potential we have
\begin{eqnarray}
E_\mu^\alpha = (\mu+1+2 \alpha) \hbar \omega \ .
\end{eqnarray}
So from the radial equation for $\mu=0$ we have eigenvalues 
$\hbar \omega$, $3 \hbar \omega$, $5 \hbar \omega$, etc.
The fact that we have a seemingly accidental degeneracy between
different representations (i. e. between different values of $\mu$)
is the result of the U2 symmetry of the Hamiltonian mentioned above.
A consequence of this symmetry
is that for a harmonic potential the total energy
depends only on the total number of phonon excitations.  This
symmetry is distinct from the circular symmetry in $x$-$y$ space.

\subsubsection{Anharmonic Potential}

The U2 symmetry is broken by anharmonic terms which then take us into
the generic case of a particle in a circularly symmetric potential
which is not harmonic.  Accordingly, we now consider the effect of
adding an anharmonic perturbation of the form $\gamma r^4$
to the harmonic potential.  For illustrative purposes,
we treat this anharmonic perturbation within first-order
perturbation theory.  Our results are characteristic of the
generic case, for which
different values of $m$ give rise to distinct eigenvalues.
In this case, the $n$-fold degenerate manifold which has energy
$n \hbar \omega$ for the harmonic potential is split into 
doublets (corresponding to the degeneracy between $+m$ and $-m$)
and, if $n$ is odd, a singlet from $m=0$.  Our explicit
results are given in Table I.
These results are generic in the sense that addition of further
anharmonic terms will not further change the degeneracies.
 
\subsection{Toy Model of Translation-Rotation Coupling}

In this section we explore the consequences of allowing coupling
between rotations and translations. Since we now
restrict attention to the manifold of $(J=1)$,
we need keep only terms with $L=2$ in Eq. (\ref{VIBROT}).  Thus,
as a toy model, we set
\begin{eqnarray}
V({\bf r}, \Omega) &=& \frac{1}{2} k r^2 
- \frac{5}{2} \alpha (3 \cos^2 \theta-1)
- \frac{5}{2} \beta r^2 \sin^2 \theta \cos ( 2 \phi - 2 \phi_r) \ ,
\end{eqnarray}
where $\alpha$, $\beta$, and $k$ are constants and
the factor $-\frac{5}{2}$ is included so that the matrix
elements are numerically simple.  Also we take the dependence on
$r$ to be quadratic to facilitate calculation of the matrix elements.
In the language of Eq. (\ref{VIBROT}) this model has
$v_2^0(r)=- \alpha \sqrt {20 \pi}$ and
$v_2^{\pm 2}(r)=- \beta r^2 \sqrt{10 \pi /3}$.  We are going to consider
the effect of this Hamiltonian within the manifold of $(J=1)$ states.
Using this toy model we can illustrate how the rotation-translation
affects the symmetry of the energy levels.
Within the $(J=1)$ manifold we may use operator equivalents to write
\begin{eqnarray}
V({\bf r}, \Omega) &=& \frac{1}{2} k r^2 
+ \alpha (3J_z^2-2) + \beta [ (J_x^2 - J_y^2)(x^2-y^2)
+ 2(J_xJ_y+J_yJ_x) xy ] \nonumber \\
&=& \frac{1}{2} k r^2 
+ \alpha (3J_z^2-2) + \frac{1}{2} \beta (J_+^2 + J_-^2)(x^2-y^2)
- i \beta (J_+^2 - J_-^2 ) xy \ .
\label{VIBJ} \end{eqnarray}
For illustrative purposes we will assume that $\alpha$
and $\beta \sigma^2$ are small compared to the phonon energy
$\hbar \omega$. 
Here $\sigma^2 = \langle x^2 \rangle = \langle y^2 \rangle$, where
the averages are taken in the ground state.
In that case, in addition to the quantum number
$P$, the total number of phonons, $N$, is a good quantum number.
However, we emphasize that in our numerical work\cite{numerics}, we do not
make this approximation.  Equation (\ref{EQN}) assumes that
$J$ is a good quantum number but mixes states with different
numbers of phonon excitations.  We should mention that the toy
model assumes that the molecule has minimal potential energy
when it on the axis of the tube.  For small [e. g. (9,0) tubes]
this assumption is justified.  For larger tubes, the minimal
potential energy occurs for a nonzero value of $r$ and the molecule
is dominantly off center.  We will later treat that case using a different
model.
 
\subsection{Results of the Toy Model}

We now discuss the results of the toy model assuming that the
number of phonons is a good quantum number. We note that all the
energy expressions given below are with respect to 
$BJ(J+1)$ with $J=1$.

\subsubsection{Zero Phonon Manifold}

We first consider the manifold of the states having $J=1$ with zero
phonons.  One finds that the energy is diagonal in $J_z$ with
\begin{eqnarray}
E(J_z) = \alpha (3J_z^2 - 2) \ ,
\end{eqnarray}
so that (if $\alpha$ is positive) 
one has the singlet $J_z=0$ state lower than
the doublet $J_z= \pm 1$ states by an energy separation of $3 \alpha$.
One may visualize this as the energy difference between a state
for which the  molecule is in the phonon ground state and is oriented
parallel to the axis and the two states when the molecule is in the
phonon ground state and is oriented transversely to the axis.
For later use we tabulate these wavefunctions in Table \ref{VIBROT0}.

\subsubsection{One-Phonon Manifold without Rotation-Translation
Coupling}

If we set $\beta=0$ in the toy model of Eq. (\ref{VIBJ}),
then essentially we have independent oscillation of molecules
which have fixed orientation.  Then if $n_x$ and $n_y$ are the
vibrational quantum numbers, we see that in the one-phonon
manifold ($n_x+n_y=1$) we have
\begin{eqnarray}
E(n_x, n_y, J_z) = 2 \hbar \omega + \alpha (3J_z^2 -2 ) \ ,
\end{eqnarray}
so that the lowest energy  state (if $\alpha>0$) is doubly degenerate and
the excited state is four-fold degenerate, as is shown in Fig.
\ref{VR1FIG}.

\subsubsection{One-Phonon Manifold with Rotation-Translation
Coupling}

The unphysical aspect of the energy level scheme we just found for
the one-phonon manifold is that it does not take into account
that the molecular orientation ought to be correlated with the
translational motion.  If the molecule translates near the wall,
then the molecule should preferentially be parallel to the wall.
This means that the orientation of the molecule has to be 
correlated with the translational motion.  This effect will be
greater the more strongly the wall potential affects the motion
of the molecule.

In terms of number operators $n_x$ and $n_y$ which are the
number of phonon excitations in the $x$-direction and $y$-direction,
respectively and $a_x^\dagger$ and $a_y^\dagger$ which are creation
operators for these excitations, we may write the Hamiltonian
for the one-phonon $(J=1)$ manifold as
\begin{eqnarray}
{\cal H} &=& \hbar \omega (n_x+n_y+1)
+ \alpha (3J_z^2-2)   \nonumber\\
&+& \sigma^2 \beta \Biggl[ (J_+^2 + J_-^2)(n_x-n_y)
- i (J_+^2 - J_-^2 )(a_x^\dagger a_y + a_y^\dagger a_x)
\Biggr] \ .
\label{VR1EQ} 
\end{eqnarray}
This gives the energy level scheme shown in the rightmost panel
of Fig. \ref{VR1FIG}.
The wavefunctions are given in Table \ref{VIBROT1} and we discuss them now.
First of all, in a classical picture, we would
argue that the molecule can oscillate equivalently in each of the
two coordinate directions transverse to the cylinder.  In each of
these two cases
the molecule can assume three inequivalent orientations because
the directions a) along the axis of the tube, b) parallel to
the directions of spatial oscillation, and c) transverse to the
direction of spatial oscillation are all inequivalent to one another.
This argument predicts that the six states form three
doubly degenerate energy levels.  
Quantum mechanically the situation is different.
In Fig. \ref{VR1FIG} we show the energy levels when no dynamical
mixing between rotations and translations is allowed, i. e.
for $\beta=0$.  In this limit
the six states form a degenerate doublet and a degenerate quartet.
When translation-rotation mixing is allowed, {\it i. e.} for
$\beta \not= 0$, we now have the generic case of two doublets
and two singlets, as shown in Fig.~\ref{VR1FIG}.  
The wavefunctions are shown in Fig.~\ref{TRANSROT2}.  
(It is interesting to note that it is not
obvious that the wave functions for $P=+2$ and $P=-2$
are related by symmetry.)

\subsubsection{Two-Phonon Manifold with Rotation-Translation
Coupling}

Actually, because the dependence on $r$ of the matrix elements in
Eq.~(\ref{VIBJ}) was taken to be either constant or proportional
to $r^2$, the representation of Eq.~(\ref{VR1EQ}) is valid within
any manifold of fixed total number of phonons and $(J=1)$.
The removal of the degeneracy  in the energy level scheme of
the two-phonon (i.e. $n_x + n_y = 2$) and $(J=1)$ manifold 
according to the Hamiltonian of  Eq.~(\ref{VR1EQ})
is shown in Fig.~\ref{n2j1levels}.
The eigenfunctions and eigenvalues for this manifold including
anharmonicity are listed in Table~\ref{ANVIBROT}.

The results obtained by numerically solving the eigenvalue problem
of  Eq.~(\ref{EQN}) using WS77 potential are given
in the last column of Table~\ref{9,0}\cite{numerics}.  
To understand the meaning of this spectrum,
we relate these results to those of the toy model when
the parameters of the toy model are suitably chosen.
For a good fit we allow the constants $\alpha$ and $\beta$
to depend on  the total number of phonons, $N$.  (This
dependence reflects the fact that the dependence of the
parameters of the toy model on $r$ is arbitrary and unrealistic.)  
In this simple model we also include the anharmonic term
$\gamma r^4$ which we treat within first order perturbation theory.
We determine the best parameters for the toy model by making
a least squares fit of the numerically determined energy levels
to those of the toy model and these parameters as well as the
 results of this fit are given in Table. \ref{9,0}.  The fact that
$\alpha_N$ depends on $N$ indicates that we should probably replace
$\alpha$ by $\alpha r^2$.  Also the fact that the splitting of the two phonon
manifold is not perfectly reproduced by the toy model indicates that the
anharmonicity energy is not simply proportional to $r^4$.
Nevertheless the close agreement between our numerical results and those
of the toy model indicates that this model provides a useful simple
picture of translation-rotation coupling.

\subsubsection{Summary}

We can summarize the systematics of the rotation-translation spectrum
of the toy model we have introduced. 
We first consider the harmonic $\gamma=0$
case and then discuss the effect of introducing anharmonicity.
In the $N$-phonon sector the harmonic phonon wavefunctions
give rise to states proportional to $(x+iy)^N$.  Combining these
with a $J_z=1$ state gives a unique
$P=N+1$ state.  This state will be degenerate with the
similar $P=-N-1$ state.  In the toy model these states have
energy $(N+1)\hbar \omega + \alpha$.
Adding anharmonicity shifts the energy of
these two states, but their degeneracy is generic.

Harmonic phonon states which transform like $(x+iy)^{N-2k}$ will
uniquely combine with $J_z=0$ states to form states for which
$P=(N-2k)$ and which have energy $(N+1)\hbar \omega - 2 \alpha$.
In analogy with Fig.~4, anharmonicity splits these states into doublets of
$+P$ and $-P$ and, if $P$ is even, a singlet for $P=0$.

The rotation-translation coupling (proportional to $\beta$)
influences the states with $P=N-1$, $P=N-3$, etc.  For positive $P$
one has two eigenstates made from linear combinations of states 
of the form $\phi_1 \equiv (x+iy)^{P+1} |J_z=-1\rangle$ and
$\phi_2 \equiv (x+iy)^{P-1} |J_z=+1\rangle$. 
Since the rotation translation coupling interaction
proportional to $xy (J_+^2 - J_-^2)$ has matrix elements
between these two states, the eigenstates
$\phi_1\pm \phi_2$ will be split by an amount proportional to
$\beta \sigma^2$
and this splitting will be modified by anharmonicity.  
Obviously, this scenario indicates that one
can not understand the degeneracies of the states of a hydrogen
molecule in confined geometry without considering the
effect of rotation-translation coupling.

\section{MEXICAN HAT POTENTIAL}

Here we discuss the case when the minimum of the potential $V_0(r)$
occurs for non-zero $r$ as happens for H$_2$ molecules inside $10 \times 10$ tubes
or for H$_2$ molecule in a bound state outside any tube.
We start from Eq. (\ref{EQN}).  To see what this equation yields,
we first consider its solutions for a $(J=0)$ molecule. 
We have solved the eigenvalue problem of
Eq. (\ref{EQN}) numerically on a mesh of points  for
a $10 \times 10$ tube\cite{numerics}. 
The  results shown in Fig.~\ref{10x10}a indicate
two different regimes for the dependence of the energy
levels on the quantum number $P$. For the low-lying energy states
it is quadratic and then gradually becomes linear as the energy
of the states increase.

It is possible to understand the quadratic behavior of energy levels versus
the quantum number $P$  based on a simple
idealized  model. Assuming $V_0(r)$ can be replaced by a harmonic  oscillator
potential, Eq. (\ref{EQN}) becomes essentially
\begin{eqnarray}
\Biggl[ - {\hbar^2 \over 2m} {\partial^2 \over \partial r^2}
+ {\hbar^2 P^2 \over 2mr^2} + E_0 +
\frac{1}{2} k (r-r_0)^2 \Biggr] f_P^{(\alpha)}(r) =
E_P^{(\alpha)} f_P^{(\alpha)}(r) \ .
\end{eqnarray}
In writing this result we dropped the term linear in the derivative.
This term does not contribute to the energy in first-order perturbation
theory.  When we treat the term in $P^2$ perturbatively, this
equation leads to a harmonic oscillator spectrum with
\begin{eqnarray}
E_P^{(N)}=E_0 + (N+ \frac{1}{2}) \hbar \omega + {\hbar^2 P^2 \over 2m}
\Biggl\langle {1 \over r^2} \Biggr\rangle \ ,
\label{PSQEQ} \end{eqnarray}
where $\omega = \sqrt{k/m}$ and
$\langle X \rangle$ here indicates an average of $X$ over the
radial wavefunction.  

The gray solid lines in Fig.~\ref{10x10}a shows the results
based on this model and the points are from the numerical 
exact results, indicating that our idealized model successfully
describes the  low-lying energy spectrum. 
Here
$\hbar \omega$ is of order 14 meV and and the quantum of tangential
kinetic energy $\langle \hbar^2/(2mr^2) \rangle$ is about 0.1 meV.
Curiously, this spectrum is reminiscent of the vibration-rotation
spectrum of a diatomic molecule\cite{HERTZ}.
Finally as we go away from the ground state, the simple model 
is not enough to explain the observed behavior. 
We note that the spacing
between the energy levels is not constant (probably due
to an anharmonic contribution to the potential) and the 
dependence on the $P$ becomes almost linear.

We next discuss the solution of
Eq. (\ref{EQN}) for a $(J=1)$ H$_{2}$ molecule.
Fig.~\ref{10x10}b shows the results
obtained numerically.\cite{numerics}. We will interprate numerical results
using a simple model which includes the  translation-rotation 
coupling (as embodied by the $v_L^M$'s). We expect the
radial wavefunctions to be
Gaussians centered about $r=r_0$.  Indeed in the terms containing
$v_L^M(r)$, we will make the replacement
\begin{eqnarray}
v_L^M(r) \rightarrow \langle v_L^M(r) \rangle \ .
\end{eqnarray}
(To a good approximation these averages can be calculated from the $(J=0)$
wavefunction.) In what follows we set
\begin{eqnarray}
\langle v_2^M (r) \rangle = w_M 
\end{eqnarray}
for $M=0$ and $M=2$.  
Then, if we set $r=r_0+x$, Eq. (\ref{EQN}) may be written as
\begin{eqnarray}
 \Biggl[ &-& {\hbar^2 \over 2m} {\partial^2 \over \partial r^2}
+ {\hbar^2 P^2 \over 2mr^2} + E_0 +
\frac{1}{2} k (r-r_0)^2 \Biggr] f_P^{(\alpha)}(r) \nonumber\\
&-& {1 \over \sqrt{2 \pi}} \sum_{M'} C(121;M-M',M') w_{M'} f_{M-M',P}^{(\alpha)}(x) 
= E_P^{\alpha} f_{M,P}^{(\alpha)} (x) \ ,
\label{EQNA} 
\end{eqnarray}
where the  Clebsch-Gordan coefficients assume the values
$C(121;L,0)=(3L^2-2)/\sqrt{10}$ and $C(121;-1,2)=\sqrt{3/5}$.
For each value of $N$ (the number of radial phonons) and $P$ (the number
of tangential excitations) the Hamiltonian is the following three dimensional matrix
(where the rows correspond to $J_z=-1$, $J_z=0$, and $J_z=+1$, in that order):
\begin{eqnarray}
{\cal H}(N,P) &=& [E_0+ N \hbar \omega] {\cal I} + \left[ \begin{array} {c c c}
\frac{1}{3} \delta + A (P-1)^2 & 0 & B \\
0 & - \frac{2}{3} \delta + AP^2 & 0 \\
B & 0 & \frac{1}{3} \delta + A (P+1)^2 \\
\end{array} \right] \ ,
\label{SIMPLE} 
\end{eqnarray}
where ${\cal I}$ is the unit matrix, $A=\langle \hbar^2/(2mr^2) \rangle$,
$\delta=-3w_0/\sqrt{20\pi}$, and $B=-w_2\sqrt{3/10\pi}$.
For fixed values of $N$ and $P$ we have the three energy eigenvalues
\begin{eqnarray}
E^{(0)} &=& E_0 + N \hbar \omega - \frac{2}{3} \delta + AP^2 \nonumber \\
E^{(\pm)} &=& E_0 + N \hbar \omega + \frac{1}{3} \delta + A(P^2+1)
\pm \sqrt{ 4A^2P^2 + B^2 } \ .
\label{FITEQ} \end{eqnarray}

In Fig.~\ref{10x10}b  we show the spectrum of a $(J=1)$ molecule obtained
numerically from Eq.~(\ref{EQN}) as a function of $P$.  Our numerical
results indicate the phonon number $N$ is a good quantum number
and can so be identified only for $N<3$.  Accordingly, we limit our
detailed interpretation in terms of the model of Eq.~(\ref{SIMPLE})
to $N=0$.  
For each value of $P$ there are three energy eigenvalues,
two of which are close in energy. These corresponds to the case where 
the H$_{2}$
molecule is oriented parallel to the tube surface (i.e. t and p orientations
in Fig.~3). The third energy corresponds to the orientation perpendicular to
the tube surface (i.e. $r$ radial orientation shown in Fig.~3). 
This orientation has an energy about 3 meV near than that of 
the other two orientations, and  is comparable to
2.6 meV observed for H$_{2}$ on graphite\cite{novaco}. 
Fig.~\ref{10x10}b  shows also the result of the simple model of 
Eq.~(\ref{SIMPLE}).  
The values of $A$, $B$, and $\delta$ used to get a good fit are given in the
figure.  The value of $A$ (0.092 meV) is not very different from the value
$\hbar^2/(2mr_0^2)=0.087$ one gets from the value of $r=r_0=3.46$~\AA 
~ at the
minimum of the potential.  Thus the numerical results are easily understood
in terms of our simple toy model.

\section{EXPERIMENTAL OBSERVATION OF THE ENERGY SPECTRUM}

Here we make some remarks concerning the observation of these modes
via inelastic neutron scattering.  Specifically we consider the
energy loss spectrum in the neutron time-of-flight spectrum.
(This technique has been used to probe local excitation of
H$_2$ molecules in the octahedral sites of C$_{60}$.\cite{NIST})
We start be recalling the results for the cross section for
inelastic neutron scattering of H$_2$ molecules. When the very
small coherent (i. e. nuclear spin independent) scattering is
neglected, the result is
\begin{eqnarray}
{\partial^2 \sigma \over \partial \Omega \partial E} &=& {k' \over k}
\Biggl[ Nx {\cal S}_{1 \rightarrow 1} + Nx {\cal S}_{1 \rightarrow 0}
+ N(1-x) {\cal S}_{0 \rightarrow 1} \Biggr] \ ,
\end{eqnarray}
where ${\bf k}$ (${\bf k}'$) is the wavevector of the incident
(scattered) neutron, $N$ is the total number of H$_2$ molecules
in the target,  $x$ is the fraction of H$_2$ molecules which
have odd $J$ (i. e. are ortho molecules), and the subscripts
indicate the partial cross sections due to ortho molecules, to
ortho-para conversion, and to para-ortho conversion, respectively.
When the sum over nuclear spin states is performed, these partial
cross sections are given by
\begin{eqnarray}
{\cal S}_{0 \rightarrow 1 , j} & = & \frac{3}{4}
(b')^2 \sum_{J_i=0,J_f=1} P_i \delta(E - E_i + E_f)
|\langle f | e^{i \kappav \cdot {\bf R}_j}
\sin( \frac{1}{2} \kappav \cdot \rhov) | i \rangle|^2  \nonumber \\
{\cal S}_{1 \rightarrow 0 , j} & = & \frac{1}{4}
(b')^2 \sum_{J_i=1,J_f=0} P_i \delta(E - E_i + E_f)
|\langle f | e^{i \kappav \cdot {\bf R}_j}
\sin( \frac{1}{2} \kappav \cdot \rhov) | i \rangle|^2  \nonumber \\
{\cal S}_{1 \rightarrow 1 , j} & = & \frac{1}{2}
(b')^2 \sum_{J_i=1,J_f=1} P_i \delta(E - E_i + E_f)
|\langle f | e^{i \kappav \cdot {\bf R}_j}
\cos( \frac{1}{2} \kappav \cdot \rhov) | i \rangle|^2  \ ,
\label{CROSS} \end{eqnarray}
where $\kappav = {\bf k}' - {\bf k}$, $R_j$ is the
center of mass of the $j$th H$_2$ molecule, and $\rhov$ is the
vector displacement of one proton relative to the other proton in
the H$_2$ molecule.  To deal with molecular orientations for
molecules where $J$ is at most unity, we write
\begin{eqnarray}
\cos (\frac{1}{2} \kappav \cdot \hat \rhov) & = & j_0(\frac{1}{2} \kappa \rho)
- 4 \pi j_2(\frac{1}{2} \kappa \rho) \sum_\mu Y_2^\mu( \hat \kappav)
Y_2^\mu(\hat \rhov)^* \nonumber \\ 
\sin (\frac{1}{2} \kappav \cdot \hat \rhov) &=& 4 \pi j_1(\frac{1}{2} \kappa \rho)
\sum_\mu Y_1^\mu( \hat \kappav)^* Y_1^\mu(\hat \rhov) \ ,
\end{eqnarray}
where $j_n$ is a spherical Bessel function.
We will assume that $\kappa$ is small enough that the term in $j_2$ can be
neglected.  Since it is not trivial to obtain a meaningful result which
properly contains the Debye-Waller factor, we proceed simply, as follows.
From the numerical solution on a mesh of points we obtain the family of
wave functions (each one denoted $|J,P;\alpha\rangle$), for which $J$ and $P$ are
good quantum numbers and $\alpha=1, 2, 3, \dots$.  (In limiting cases one
may replace the single index $\alpha$ by two indices $N$ and $\gamma$, where
$N$, the number of phonons, is nearly a good quantum number.)
If we label the radial mesh points by $k=1, 2, 3 ...$, then we have
\begin{eqnarray}
|J,P; \alpha\rangle &=&
{\sum_{\mu=-J}^J \sum_k c_{J,P}^{(\alpha)} (k, \mu) e^{-i (P+\mu) \phi_r}
\sqrt {r_k} |r_k \rangle |J,J_z=\mu\rangle \over
\Biggl[ 2 \pi  \sum_{\mu=-J}^J \sum_k |c_{J,P}^{(\alpha)}(k,\mu) |^2
r_k \Biggr]^{1/2}} \ .
\end{eqnarray}
Here $|r_k\rangle$ (and later $\langle r_k | $) 
is a wavefunction of unit amplitude at the position $r_k$.
Also the $c_{J,P}^{(\alpha)}(k,\mu)$'s are the set of coefficients (for
fixed $J$, $P$, and $\alpha$) which are obtained by the numerical
solution of the $(2J+1)$-component radial eigenvalue problem on a set
of mesh points $\{r_k\}$.  This discretized eigenvalue problem involves
diagonalization of nonsymmetric matrix.  (The radial
equation gives rise to a Hermitian problem only if proper
account is taken of the radial weight function.)  The numerical
program takes no account of any weight factor, but rather normalizes
these wavefunctions by requiring that the sum of the squares of their
coefficients be unity. Since we always wish to define inner
products with a weight factor $r_k$, 
we will explicitly include a factor $r_k$
when we take inner products.  
Then if $X$ is a quantity which is local in $r$ and $\phi_r$
but may be off-diagonal in $J$ and/or $J_z$, we express its
matrix element between such numerically obtained wavefunctions as
\begin{eqnarray}
&& \langle J', P';\beta| X | J,P;\alpha \rangle \nonumber \\
&\equiv & { \int_0^{2 \pi} d \phi_r \sum_{k,\mu, \mu'} 
\langle J', J_z=\mu'| X(r_k, \phi_r)| J, J_z =\mu \rangle r_k
c_{J',P'}^{(\beta)}(k,\mu')^* c_{J,P}^{(\alpha)}(k,\mu)
e^{i(P'-P+\mu'-\mu)\phi_r} 
\over 2 \pi \Biggl[ \sum_{k,\mu} |c_{J,P}^{(\alpha)} (k,\mu)|^2 r_k
\sum_{k',\mu'} |c_{J',P'}^{(\beta)} (k',\mu')|^2 r_{k'} \Biggr]^{1/2} } \ .
\end{eqnarray}
Here $\mu'$ assumes integer values between $-J'$ and $+J'$ and
$\mu$ integer values between $-J$ and $+J$. 

To evaluate the cross sections, the major problem is to evaluate
the matrix element, which we may call $<f|X|i>$.  We will not
discuss all possible transitions (which are shown in Fig.~4 of  I).
Instead we will focus on
the neutron energy loss spectrum due to a) para to ortho conversion
and b) radial phonon creation on an ortho molecule.

\subsection{Para to Ortho Conversion}

For  neutron energy loss due to para to ortho
conversion, we need the matrix element 
\begin{eqnarray}
X&=&4 \pi j_1(\frac{1}{2} \kappa \rho) \exp(i \kappav \cdot {\bf R}_j) \sum_\nu
Y_1^\nu(\kappav) Y_1^\nu(\hat \rhov)^*
\end{eqnarray}
and for simplicity we consider the case when $\kappav$ is perpendicular
to the cylindrical axis of the nanotube.  Then we may as well place
$\kappav$ along the local $x$ axis.  Also the center of mass of
molecule $j$ is at ${\bf r}_k$ relative to the axis of the tube
at position ${\bf R}_j^{(0)}$ which contains the $j$th molecule.
Then we may write
\begin{eqnarray}
X &=& \sqrt{6 \pi} j_1(\frac{1}{2} \kappa \rho)
e^{i \kappav \cdot {\bf R}_j^{(0)}}
e^{i \kappa r_k \cos \phi_r} [Y_1^{-1}(\hat \rhov) - Y_1^1(\hat \rhov)] \ .
\end{eqnarray}
At low temperature the initial state (whose energy is denoted $E_{i}$)
will be the ground state for $J=0$
and for a small value of $P$ (which we denote $P_i$).  Thus
\begin{eqnarray}
S_{0 \rightarrow 1} = \frac{9}{2} \pi [b' j_1(\frac{1}{2} \kappa \rho)]^2
Z^{-1} \sum_{i,f} e^{- E_i/(kT)}
\delta[E - E_i + E_f ] M_{if} \ ,
\end{eqnarray}
where $E_f$ is the energy of the $(J=1)$ final state, and
$Z= \sum_i \exp [- E_i/ (kT)]$, and
\begin{small}
\begin{eqnarray}
M_{if} &=&  \Biggl| { \int_0^{2 \pi} d \phi_r \sum_{k,\mu}
\langle J=1, J_z=\mu| Y_1^1(\hat \rhov) - Y_1^{-1}(\hat \rhov) | J=0, J_z=0 \rangle
e^{i \kappa r_k \cos \phi_r} e^{i(P_f-P_i+\mu)\phi_r} r_k c_f(k,\mu)^* c_i(k)
\over 2 \pi \Biggl[ \sum_k |c_i (k)|^2 r_k
\sum_{k,\mu} |c_f (k,\mu)|^2 r_k \Biggr]^{1/2} } \Biggr|^2 \nonumber \\
&=& { \Biggl| \sum_k [ J_{P_f-P_i+1}(\kappa r_k)c_f(k,1)^*
- J_{P_f-P_i-1}(\kappa r_k) c_f(k,-1)^*]r_k c_i(k) \Biggr|^2
\over 4 \pi \sum_k |c_i (k)|^2 r_k
\sum_{k,\mu} |c_f (k,\mu)|^2 r_k } \ ,
\label{Mif}
\end{eqnarray}
\end{small}
where $J_n(x)$ is a Bessel function and 
$c_{i}(k)$  is the  value of  
$c_{J,P}^{(\alpha)}(k)$ for the initial state 
and $c_{f}(k)$ is  the value of 
$c_{J',P'}^{(\beta)}(k',\mu')$ for the final state.

We now discuss the qualitative meaning of this result. 
We calculated $S_{0 \rightarrow 1}$ for several temperatures and
the results are plotted in Fig.~8a.  At zero
temperature the initial state has $P_i=0$.  So the energy loss is zero
up to the cut-off energy which is the energy of para-to-ortho conversion.
For small temperatures, the cross section
does not appear discontinuously, but  turns on rapidly over
a range of energy of order $kT$. The first para-to-ortho (i.e.
$J=0$ to $J=1$) 
transition is observed at energies about 16.8 and
13.6 meV with approximately  one-to-two intensity ratio, 
(corresponding to a splitting of 3.2 meV
between $J=1, M=0$ and $J=1, M=\pm1$ states, respectively).
The center of gravity of the $J=1$ levels gives the average
para-ortho conversion energy, $E_{c} = 14.67$ meV. This
result represents  only a small amount of downward
shift of 0.03 meV from the free molecule value of $14.7$ meV.

We note that there are several neutron scattering 
experiments reporting the para-to-ortho transition\cite{brown,price}. 
The observed splitting is about 1 meV, 
suggesting the idea that in those experiments
hydrogen molecules were probably not inside the nanotubes.
The calculated para-to-ortho splitting of 3.2 meV is slightly
larger than the 2.6 meV splitting observed for  H$_{2}$ on 
graphite\cite{novaco}.

In addition to the sharp para-to-ortho rotational transitions,
Fig.~8a also indicates several  broad radial phonon transitions
at energies about 15 and 30 meV
(similar values to those of H$_{2}$ trapped in solid C$_{60}$).
Finally we note that there are many lines in the spectrum
due to transition between different tangential phonon states
(i.e. $P$ quantum number). However their observation could be
problematic due to experimental energy resolution (which
would be worse at high energies than FWHM of 0.5 meV  
used in Fig.~8).

\subsection{Ortho Cross Section}

Here we discuss the scattering from an ortho H$_2$ molecule.
As before, the major problem is the calculation of the matrix element.
In analogy with the previous results we write
\begin{eqnarray}
S_{1 \rightarrow 1} = \frac{1}{2} [b' j_0(\frac{1}{2} \kappa \rho)]^2
Z^{-1} \sum_{if} e^{-E_i/(kT)} \delta[E - E_i + E_f ] M_{if} \ ,
\end{eqnarray}
where we neglect terms involving $j_2({1 \over 2} \kappa \rho)$ and
\begin{eqnarray}
M_{if} &=& | \langle f| e^{i \kappa r_k \cos \phi_r}| i \rangle |^2 \ .
\end{eqnarray}
Now for $M_{if}$ we have in the notation of Eq.~(\ref{Mif})
\begin{eqnarray}
M_{if} &=&  \Biggl| { \int_0^{2 \pi} d \phi_r \sum_{k,\mu}
e^{i \kappa r_k \cos \phi_r} e^{i(P_f-P_i)\phi_r} r_k c_f(k,\mu)^* c_i(k,\mu)
\over 2 \pi \Biggl[ \sum_{k,\mu} |c_i (k,\mu)|^2 r_k
\sum_{k,\mu} |c_f (k,\mu)|^2 r_k \Biggr]^{1/2} } \Biggr|^2 \nonumber \\
&=& { \Biggl| \sum_{k,\mu} J_{P_f-P_i}(\kappa r_k)r_k c_f(k,\mu)^*
c_i(k,\mu) \Biggr|^2  \over \sum_{k,\mu} |c_i (k,\mu)|^2 r_k
\sum_{k,\mu} |c_f (k,\mu)|^2 r_k } \ .
\label{CS} \end{eqnarray}

Figure~8b shows the calculated spectrum ortho cross section
$S_{1 \rightarrow 1}$, indicating many transitions between
a large number of states. It is possible to identify the
radial phonon transitions for only one and two phonon states
as indicated in the figure.
On the other hand, the transitions between tangential phonon states
(i.e. states with different quantum number $P$) dominate the
calculated spectrum, giving rise to many sharp peaks. 
Due to experimental energy resolution, it is
probably not possible to observe the transition at high energies
(say above 20 meV). However the resolution at energies below
around 10 meV could be about 0.5 meV (which is used in Fig.~8)
and therefore it may be possible to observe these transitions. 

Finally we note that the tangential phonon transitions
below 10 meV show a maximum near 3.6 to 4 meV.  We can understand
this by considering the condition that the neutron waveform
become resonant with the wavefunction of H$_2$ molecule going
around the circumference of the minimum of the Mexican hat.
The phase change when the neutron passes through a diamater
of the Maxican hat is $2\kappa r_0$, where $r_0$ is the
radius at which the Maxican hat potential is minimal.
The phase change of the H$_2$ molecule going around half
a circumference is $\pi P$.  If we assume an initial state
with $P_i=0$, then the resonance condition is
$\pi P_f = 2\kappa r_0$.  With $\kappa=3 \AA^{-1}$
and $r_0= 3.5 \AA$, we find $P_f \approx 6$.
Then $E_f-E_i = \hbar^2 P_f^2 \langle (2mr_0)^{-1}\rangle=
0.09P^2 $ meV $= 3.2$ meV, in reasonable agreement with
the numerical evaluation.

\section{Conclusion}

We list the major conclusion from our study of H$_2$ molecules
bound to nanotubes which we treat as smooth cylinders.

$\bullet$  We have derived the analog of a radial equation for
the Schrodinger equation for the translational and rotational
motion of a molecule in cylindrical geometry.  This formulation
leads to classifying translation-rotation wavefunctions
according to their properties under a global rotation of the
molecule about the cylindrical axis.

$\bullet$
Using this radial equation, the translation-rotation wavefunctions
for a hydrogen molecule bound either inside or outside a nanotube
can be obtained numerically.  We also have developed simple
toy models which quite accurately reproduce the numerical
results, but have the advantage that they elucidate the
nature of the translation-rotation dynamics.  

$\bullet$  Simple classical symmetry arguments
fail to predict the correct degeneracies of translation-rotation
wavefunctions. However, the quantum wavefunctions are 
easy to understand qualitatively.  For instance,
for a $J=1$ molecule (such as ortho-H$_2$), one class
of translation-rotation wavefunctions has
the molecule in a $J_z=0$ state (i. e. aligned along
the axis of the nanotube) with no admixture from
$J_z=\pm1$.  This simplification is a result of the
mirror plane perpendicular to axis of the cylinder.

$\bullet$ We also suggest that neutron time-of-flight spectra
could provide useful confirmation of our results.  To
that end we have calculated typical spectra that might
be observed.  These are shown in Fig.~8. 

\vspace{0.2 in} \noindent
{\bf ACKNOWLEDGMENT}
A. B. H. would like to thank NIST for its hospitality during several
visits when this work was done. We acknowledge partial support from the
Israel-US Binational Science Foundation (BSF). 

\newpage
\appendix
\section{Inclusion of Both Anharmonicity and Translation-Rotation Coupling}
Here we study the simultaneous effect of anharmonicity and translation
rotation coupling for the two-phonon manifold
within the toy model.  From Table \ref{TABANH} we see that we may
write the anharmonic Hamiltonian, $V_{AH}$, which is independent of $J_z$, as
\begin{eqnarray}
V_{AH} = V_0 {\cal I} + \frac{8}{3} \gamma \sigma^4 \Biggl(
2 |2, m= 0 \rangle \langle 2,m=0 |  
- |2, m= 2 \rangle \langle 2, m=2 |  
- |2, m= -2 \rangle \langle 2, m=-2 | \Biggr)  \ ,
\end{eqnarray}
where $V_0 = 152 \gamma \sigma^4/3$
and $|2, m\rangle$ is a two-phonon wavefunction
(with energy $3 \hbar \omega$) as given in Table \ref{TABANH}.
The translation rotation interaction may be written as
\begin{eqnarray}
V &=& \alpha [ 3 J_z^2 -2] + V' \end{eqnarray}
where
\begin{eqnarray}
V' &=& \frac{1}{2} \beta \Biggl(
J_+^2 (x-iy)^2 + J_-^2 (x+iy)^2 \Biggr) \ .
\end{eqnarray}

If we write the ground state wave function as
\begin{eqnarray}
|0 \rangle = N_{0,0} e^{- \frac{1}{4} (x^2+y^2)/\sigma^2} \ ,
\end{eqnarray}
then the two phonon eigenstates are
\begin{mathletters}
\begin{eqnarray}
|2, m=0 \rangle &=& \frac{1}{2} (X^2 + Y^2 - 2) N_{00}
e^{-\frac{1}{4} (X^2+Y^2)}
\end{eqnarray}
and
\begin{eqnarray}
|2, m=\pm 2 \rangle &=& \frac{1}{2\sqrt 2} (X + iY)^2  N_{00}
e^{-\frac{1}{4} (X^2+Y^2)} \ .
\end{eqnarray}
\end{mathletters}
where $X=x/\sigma$ and $Y=y/\sigma$.  If
$\langle \ \ \rangle$ indicates a ground state average,
then we have the evaluation
\begin{eqnarray}
\langle 0 | (x-iy)^2 | 2; m=2 \rangle &=&
\sigma^2 \left \langle \Biggl( \frac{1}{2} (X^2+Y^2-2) (X-iY)^2
{1 \over 2 \sqrt 2} (X+iY)^2 \Biggr) \right \rangle \nonumber \\
&=& 4 \sqrt 2 \sigma^2 \ ,
\end{eqnarray}
so that
\begin{eqnarray}
V' &=& 2 \sqrt 2 \beta \sigma^2 \Biggl(
|2; m=-2 \rangle \langle 2;m=0 | + |2;m=0 \rangle \langle 2; m=2 |
\Biggr) J_+^2 + {\rm H. \ c. } \ , 
\end{eqnarray}
where H. c. indicates the Hermitian conjugate of the preceding term.
We find the eigenvalues to be those of Table \ref{ANVIBROT}.

\section{Cartesian Representation}

Here we rewrite the eigenfunctions in the Cartesian representation.
Write
\begin{eqnarray}
| J_z=0 \rangle  &=& | Z \rangle \ , \nonumber \\
| J_z=1 \rangle &=& - \frac{1}{\sqrt 2} |X + iY \rangle \ , \nonumber \\
| J_z=-1 \rangle &=& \frac{1}{\sqrt 2} |X-iY \rangle \ .
\end{eqnarray}
Look at Table \ref{VIBROT1}.  Eigenfunctions with energy
$\alpha \pm 4 \sigma^2 \beta$ are
\begin{eqnarray}
\psi_\pm &=& {1 \over 2 \sqrt 2}
\Biggl( -|(X+iY)(x-iy)\rangle \mp |(X-iY)(x+iy)\rangle \Biggr)
\end{eqnarray}
So, apart from a phase factor,
\begin{eqnarray}
\psi_+ &=&  {1 \over \sqrt 2} | xX + yY \rangle \nonumber \\ 
\psi_- &=&  {1 \over \sqrt 2} | xY - yX \rangle \ . 
\end{eqnarray}
The eigenfunctions with energy $\alpha$ are
\begin{eqnarray}
\psi_\pm &=& - \frac{1}{2} |(x\pm iy)(X\pm iY) \rangle
\end{eqnarray}
so that we may take the eigenfunctions to be
\begin{eqnarray}
\psi &=& \frac{1}{\sqrt 2} |xX- yY \rangle \nonumber \\
&=& \frac{1}{\sqrt 2} |xY + yX \rangle \ .
\end{eqnarray}

%%%%%%%%%%%%%% References %%%%%%%%%%%%%%%%%%%%%%

%%%%%%%%%%%%%% Tables %%%%%%%%%%%%%%%%%%%%%%
\newpage
% Table I
\begin{table}
\caption{Two-Dimensional Harmonic Oscillator Wave functions}
\vspace{0.2 in}
\begin{tabular} {| c | c | c || c | c | }
\hline \hline
\ \ $E/(\hbar \omega)$ \ \  & \ \ $dE/d \gamma)_{\gamma=0}$ \ \
&\ \  $m^{\rm a}$ \ \  & $\psi$ & $\psi(x,y)^{\rm b}$ \\
\hline
1 &  $8 \sigma^4$ & 0 & (0,0) & 1 \\
\hline \hline
2 & $24 \sigma^4$ & 1 & $(1,i)$ & $x+iy$ \\
2 & $24 \sigma^4$ & -1 & $(1,-i)$ & $x-iy$ \\
\hline \hline
3 & $56 \sigma^4$ & 0 & [ (2,0) + (0,2)] & $(r^2/\sigma^2) - 2$ \\
\hline
3 & $48 \sigma^4$ & 2 
& $[ (2,0) - (0,2) + i \sqrt 2 (1,1) ]$ & $(x+iy)^2/\sigma^2$ \\ 
3 & $48 \sigma^4$ & -2 
& $[ (2,0) - (0,2) - i \sqrt 2 (1,1) ]$ & $(x-iy)^2/\sigma^2$ \\ 
\hline \hline
4 & $96 \sigma^4$ & 1 
& $[ \sqrt 3 (3,0) + (1,2) ]$ & $[(x+iy)/\sigma][(r^2\sigma^2)- 4 ]$  \\
4 & $96 \sigma^4$ & -1 
& $[ \sqrt 3 (0,3) + (2,1) ]$ & $[(x-iy)/\sigma][(r^2/\sigma^2)- 4]$ \\
\hline
4 & $80 \sigma^4$ & 3 & $[(3,0) - \sqrt 3 (1,2) ] + i \sqrt 3 (2,1) -i (0,3)$
& $[(x+iy)/\sigma]^3$ \\
4 & $80 \sigma^4$ & -3 & $[(0,3) - \sqrt 3 (2,1)] - i \sqrt 3 (2,1) + i (0,3)$
& $[(x-iy)/\sigma]^3$ \\
\hline \hline
\end{tabular}
\label{TABANH}

\vspace{0.2 in} \noindent
(a)  In cylindrical  coordinates the $\phi$-dependence is through the factor
$e^{im \phi}$.

\vspace{0.2 in} \noindent
(b)  The wave function contains, in addition to the factors listed,
$\exp[-(x^2+y^2)/(4 \sigma^2)]$ as well as a normalization factor.
\end{table}

\newpage
\begin{table}
\caption{Wavefunctions for Rotation-Vibration for $J=1,n_x=n_y=0$.}
\vspace{0.2 in}
\begin{tabular} {| c || c || c | c | c ||}
\hline\hline
Energy$^{\rm a}$ & $P^{\rm b}$ & \multicolumn{3} {c||} {Wavefunction} \\
\hline
& & \multicolumn{3} {c||} {$n_x=0, n_y=0$} \\
\hline
& & $J_z=+1$ & $J_z=0$ & $J_z=-1$ \\
\hline \hline
\ \ $- 2 \alpha$ \ \  & 0 & 0 & $1$ & 0 \\
$\alpha$ & $1$ & $1$ & 0 & 0 \\
$\alpha$ & $-1$ & 0 & 0 & 1 \\
\hline\hline
\end{tabular}
\label{VIBROT0}

\vspace{0.2 in} \noindent
(a)  We tabulate the energy relative to $\hbar \omega$.

\noindent
(b) $P$ defines the transformation of the wave function under a global
rotation, as explained in connection with Eq. (\ref{PEQ}).
\end{table}

\newpage
\begin{table}
\caption{Wavefunctions for Rotation-Vibration for $J=1,n_x+n_y=1$.}
\vspace{0.2 in}
\begin{tabular} {| c || c || c | c | c || c | c | c ||}
\hline\hline
Energy$^{\rm a}$ & $P^{\rm b}$ & \multicolumn{6} {c||} {Wavefunction} \\
\hline
& & \multicolumn{3} {c||} {$n_x=1, n_y=0$} & \multicolumn{3}
{c||} {$n_x=0, n_y=1$} \\
\hline
& & $J_z=+1$ & $J_z=0$ & $J_z=-1$ & $J_z=+1$ & $J_z=0$ & $J_z=-1$ \\
\hline \hline
\ \ $\alpha + 4 \sigma^2 \beta$ \ \  & 0 & $\frac{1}{2}$
& 0 & $\frac{1}{2}$ & $- \frac{1}{2} i$ & 0 & $\frac{1}{2} i$ \\
$\alpha - 4 \sigma^2 \beta$ & $0$ & $\frac{1}{2}$ & 0 
& $- \frac{1}{2}$ & $- \frac{1}{2} i$ & 0 & $- \frac{1}{2} i$ \\
$- 2 \alpha$ & 1 & 0 & $\frac{1}{\sqrt 2}$
& 0 & 0 & $\frac{1}{\sqrt 2}i$ & 0 \\
$- 2 \alpha$ & -1 & 0 & $\frac{1}{\sqrt 2}$
& 0 & 0 & $- \frac{1}{\sqrt 2}i$ & 0 \\
$\alpha$ & 2 & $\frac{1}{\sqrt 2}$ & 0
& 0 & $\frac{1}{\sqrt 2}i$ & 0 & 0 \\
$\alpha$ & -2 & 0 & 0 & $\frac{1}{\sqrt 2}$ & $0$ &
0 & $- \frac{1}{\sqrt 2} i$ \\
\hline\hline
\end{tabular}

\vspace{0.2 in} \noindent
(a)  We tabulate the energy relative to $2\hbar \omega$.

\noindent
(b) $P$ defines the transformation of the wave function under a global
rotation, as explained in connection with Eq. (\ref{PEQ}).
\label{VIBROT1} 
\end{table}

\newpage
\begin{table}
\caption{Wavefunctions for $J=1,n_x+n_y=2$ with anharmonic
(scaled with $\gamma$)
and rotation-translation coupling (scaled with $\sigma$).}
\vspace{0.2 in}
\begin{tabular} {|c || c || c ||}
\hline\hline
\ \ $P$ \ \ &  Wavefunction$^{\rm a}$ & Energy$^{\rm b}$ \\
\hline \hline
0&  $|m=0; M=0 \rangle$ & $- 2 \alpha + 16 \gamma \sigma^4$ \\
1 & $[|m=2;M=-1 \rangle + |m=0; M=1\rangle] /\sqrt2$ &
$\alpha + 12 \gamma \sigma^4 + 4R$ \\
$-1$ & \ \ $[|m=-2;M=1\rangle + |m=0; M=-1\rangle] /\sqrt2$ &
$\alpha + 12 \gamma \sigma^4 + 4R$ \\
1 & $[|m=2;M=-1 \rangle - |m=0; M=1\rangle] /\sqrt2$ &
\ \ $\alpha + 12 \gamma \sigma^4 - 4R$ \ \ \\
$-1$ &$[|m=-2;M=1\rangle - | m=0; M=-1\rangle] /\sqrt2$ &
$\alpha + 12 \gamma \sigma^4 - 4R$ \\
2&  $|m=2; M=0 \rangle$ & $-2 \alpha + 8 \gamma \sigma^4$ \\
$-2$ &  $|m=-2; M=0 \rangle$ & $-2 \alpha + 8 \gamma \sigma^4$ \\
3 &  $|m=2; M=1 \rangle$ & $\alpha + 8 \gamma \sigma^4$ \\
$-3$ & $|m=-2; M=-1 \rangle$ & $\alpha + 8 \gamma \sigma^4$ \\
\hline\hline
\end{tabular}

\vspace{0.2 in} \noindent
a) Here $M$ indicates a wavefunction for which $J_z=M$.
The states indicated by $m$ are the phonon states in the
cylindrical gauge and are listed in Table \ref{TABANH}.

\vspace{0.2 in} \noindent
b) Here $R = \sqrt{ \gamma^2 \sigma^8 + 2 \beta^2 \sigma^4}$.
Also these are energies relative to $3 \hbar \omega + 40 \gamma \sigma^4$,
which is the energy for equally spaced levels extrapolating from the
zero phonon and one phonon level.
\label{ANVIBROT} 
\end{table}

\newpage
\begin{table}
\caption{Energy Levels in meV for the toy model for a $(J=1)$ hydrogen
molecule inside of a (9,0) tube,
compared to numerical calculations based on Eq. (\ref{EQN}):
The parameters (in meV) are $\alpha_0 = 2.82$. $\alpha_1=3.56$,
$\alpha_2=4.32$, $\beta_1 \sigma^2 = 0.59$, $\beta_2 \sigma^2=0.55$,
$\gamma \sigma^4 = 0.29$, and $\hbar \omega=27.36$.}
\begin{tabular} {| c || c | c |}
\hline\hline
 $\ \ P\ \ $  & Toy Model Energy$^{\rm a,b}$ &
Energy$^{\rm a}$ (Numeric) \\
\hline \hline
 0 & 0 & 0 \\
 $\pm 1$ & $3 \alpha_0 = 8.46$ & 8.46 \\ \hline
 0 & $\ \ 2 \alpha_0 + \alpha_1 + 4 \beta_1 \sigma^2 + \hbar\omega
= 38.92\ \ $ & $\ \ 39.01 \ \ $ \\
0 & $2 \alpha_0 + \alpha_1 - 4 \beta_1 \sigma^2 + \hbar\omega
= 34.20$ & 34.29 \\
 $\pm 1$ & $2 \alpha_0 - 2 \alpha_1 + \hbar \omega = 25.88$ & 25.89 \\
 $\pm 2$ & $2 \alpha_0 + \alpha_1 + \hbar \omega = 36.56$ & 36.56 \\
\hline
 0 & $2 \alpha_0 - 2 \alpha_2 + 4 \delta + 2 \hbar \omega = 56.36$
& 55.67 \\
 $\pm 1$ & $2 \alpha_0 + \alpha_2 + 3 \delta + 4R + 2 \hbar \omega=71.68$
& 71.13 \\
 $\pm 1$ & $2 \alpha_0 + \alpha_2 + 3 \delta - 4R + 2 \hbar \omega=64.64$
& 64.12 \\
 $\pm 2$ & $2 \alpha_0 - 2 \alpha_2  + 8 \gamma \sigma^4 + 2 \hbar \omega = 54.04$
& 53.53 \\
 $\pm 3$ & $ 2 \alpha_0 + \alpha_2 + 8 \gamma \sigma^4 + 2 \hbar \omega = 67.00$
& 66.35 \\
\hline\hline
\end{tabular}
\label{9,0} 

\vspace{0.2 in} \noindent
a) The zero of energy is taken to be the lowest $P=0$ level.

\vspace{0.2 in} \noindent
b) Here $R = \sqrt { \gamma^2\sigma^8  + 2 \beta_2^2 \sigma^4}$.
\end{table}

%%%%%%%%%%%%%%%%%%%% FIGURES %%%%%%%%%%%%%%%%%

\newpage
.
\begin{figure}
\includegraphics[scale=0.5,width=10cm]{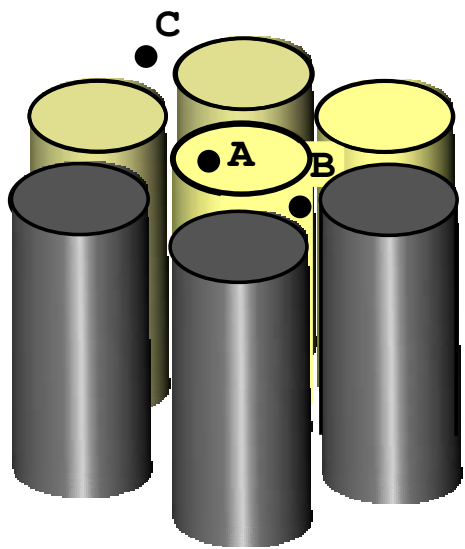}

\vspace{0.2 in}
\vfill
\caption{A schematic representation of a single wall carbon nanotube
rope indicating three different absorption sites; namely
(A) endohedral, (B) interstitial, and (C) external adsorption sites,
respectively.
}
\label{habsites}
\end{figure}

\newpage
\begin{figure}
\includegraphics[width=10cm]{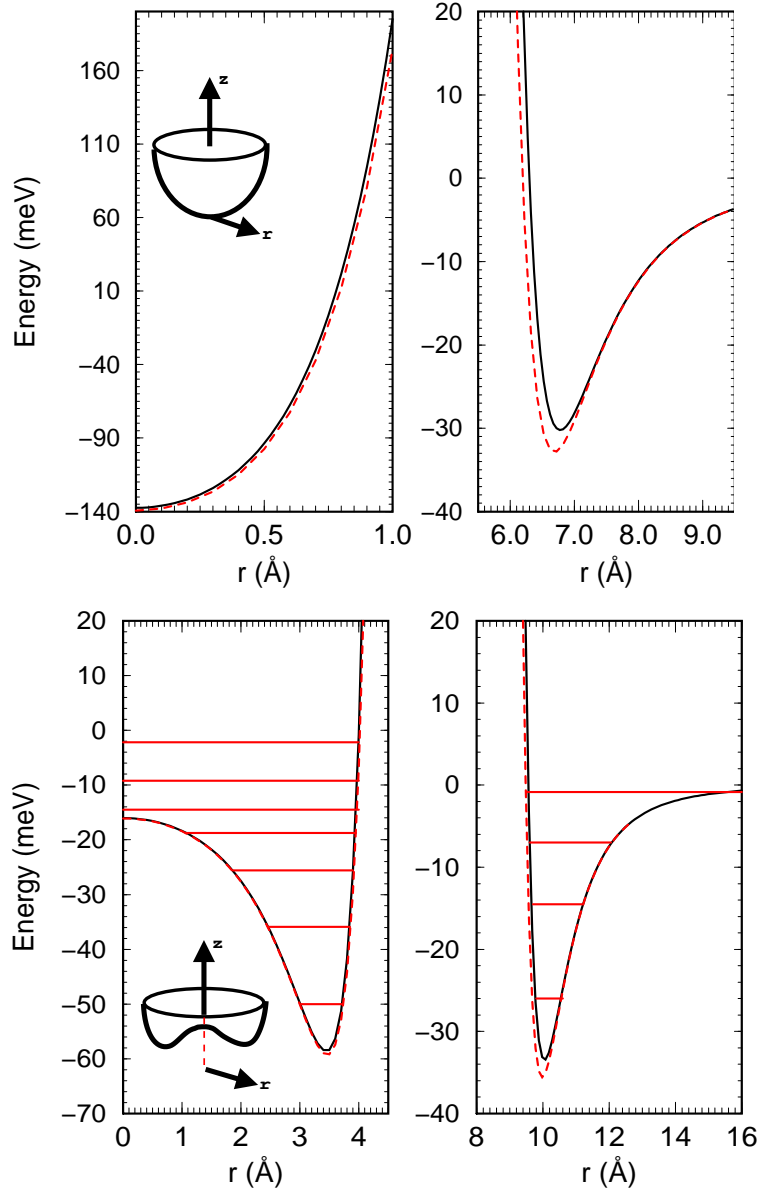}

\vfill
\caption{Potential energy as a function of distance 
from tube center for a para
hydrogen (i.e. $J=0$) interacting with  
a (9,0) (top) and (10,10) (bottom) nanotube.
The solid and dashed lines are for smooth and actual carbon
nanotubes, respectively.
For the actual carbon nanotubes, the value of z-axis
is taken arbitrarily.
The horizontal lines in the bottom panel  indicate the 
radial phonon bound states in smooth nanotubes.
The insets to the left panels are  schematic plots of
the parabolic (top) and Mexican hat potentials,
respectively.
}
\label{potvsr1}
\end{figure}

\newpage
\begin{figure}
\includegraphics[width=10cm]{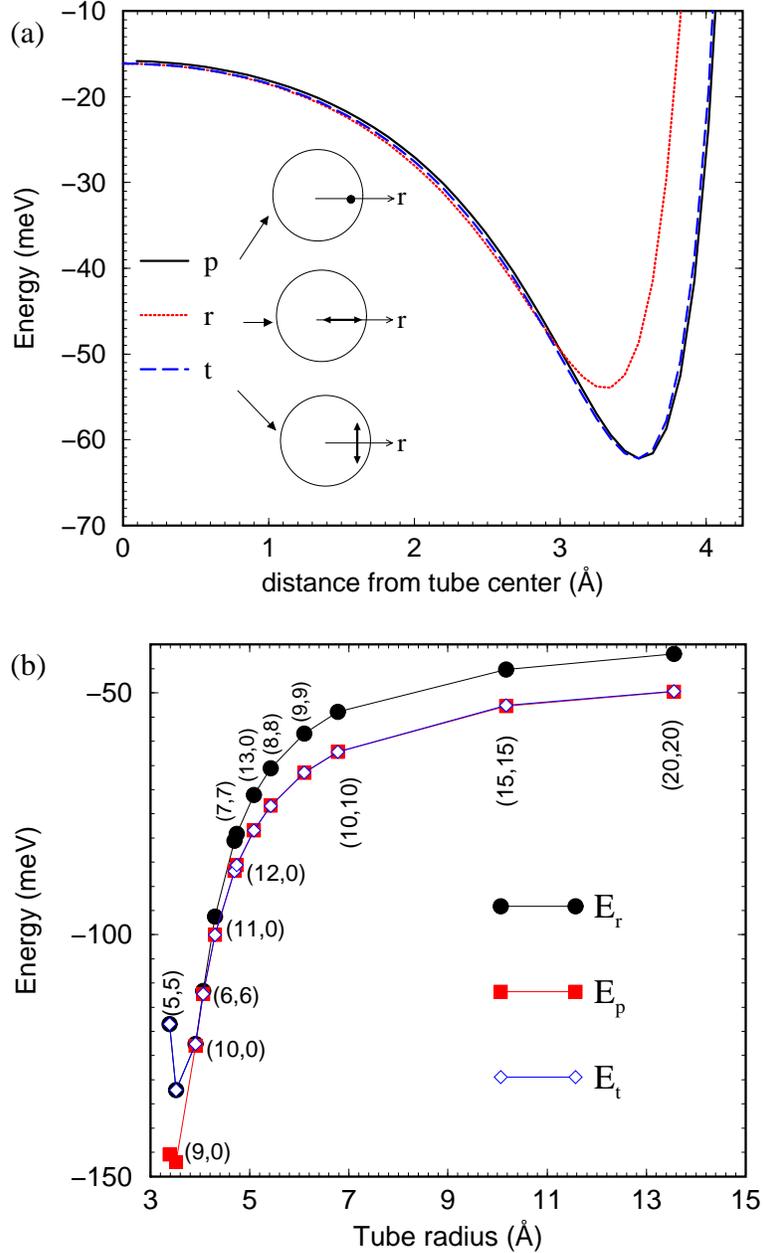}

\vfill
\caption{(a)  Potential energy as a hydrogen molecule
is translated  from the center of a (10,10) nanotube
when H$_{2}$ is oriented to be parallel to the tube axis (p),
radially (r), and tangentially (t), respectively.
Inset shows these configurations. The equilibrium distance
and minimum potential depends strongly on the orientation
of the H$_{2}$ molecule, indicating strong rational-translational
coupling. (b) The minimum potential energies, 
E$_{r}$, E$_{p}$, and E$_{t}$, for radial, parallel, 
and tangential orientation of H$_{2}$ molecule for
various nanotubes. 
}
\label{potvsr2}
\end{figure}

\newpage
\begin{figure}
\includegraphics[]{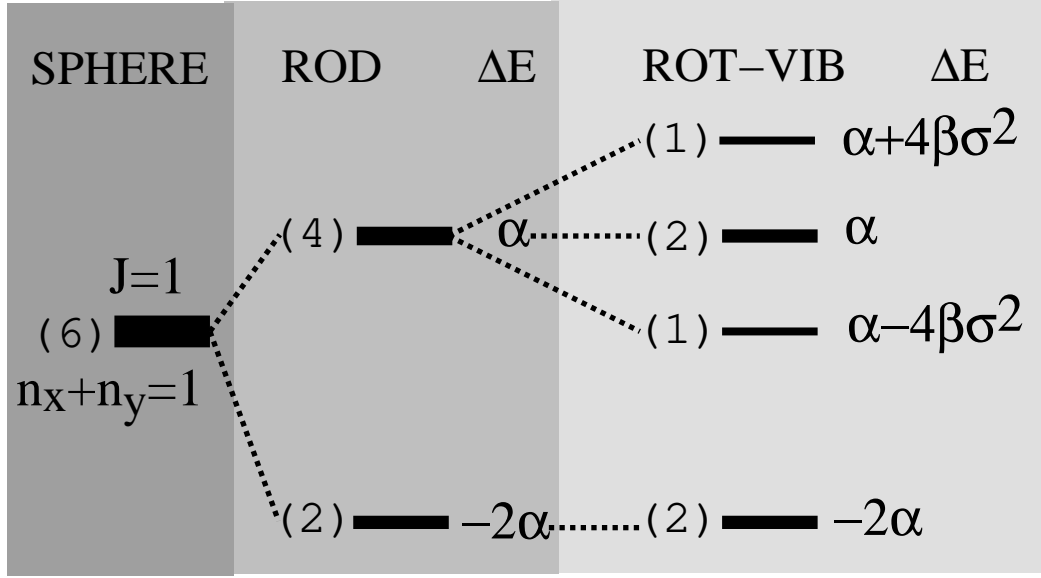}

\vspace{0.2 in} 
\vfill
\caption{Removal of the degeneracy in the energy level
scheme of the one-phonon $(J=1)$ manifold according to the Hamiltonian
of Eq. (\ref{VR1EQ}).  The diagram labeled ``SPHERE" is for a spherical
molecule for which $\alpha=\beta=0$.  That labeled ``ROD" is for
decoupled rotations and translations of a rod-like molecule
for which  $\alpha \not=0$.
That labeled ``ROT-VIB" is for translation-rotation coupling
with $\beta \not=0$.}
\label{VR1FIG}
\end{figure}

\newpage
\begin{figure}
\includegraphics[width=8cm]{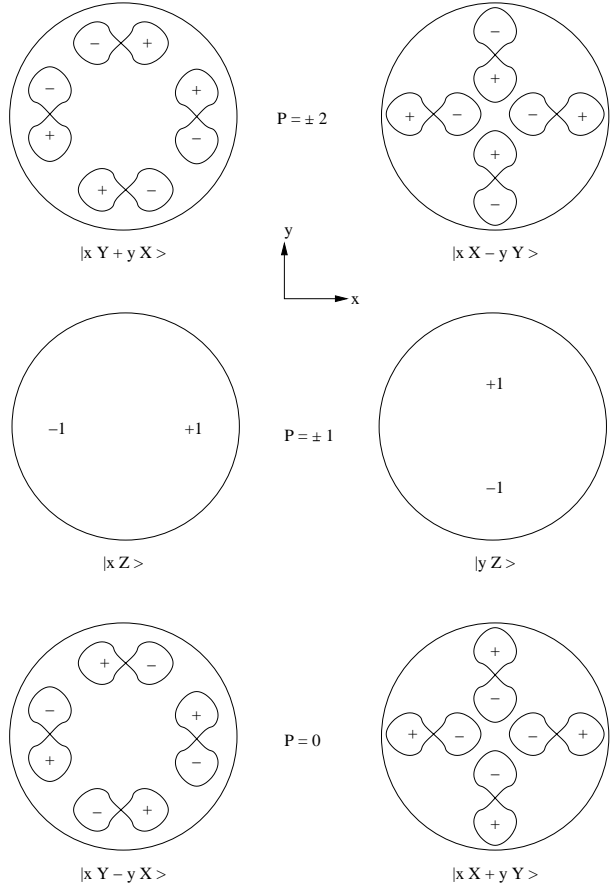}

\vfill
\caption{Translation-rotation wavefunctions for
a $(J=1)$ with one phonon when there is dynamical mixing of
translations and rotations.  Here the plane of the paper is the $x-y$
plane and each figure eight represents an $|X\rangle$ or $|Y\rangle$
orientational wavefunction and the sign associated with each lobe
of this $p$-like function is indicated. For the $|Z\rangle$ orientational
function (which would have the figure eight coming out of the page) we
indicate the sign of the lobe in front of the page.  Each
orientational wave function is multiplied by a translational
wavefunction $|x\rangle$ or $|y\rangle$, where for
instance $|x \rangle \sim x \exp[- \frac{1}{4} (x/\sigma)^2]$.
The presence of a phonon in the $r_\alpha$ coordinate
thus causes the wave function to be an odd function of
$r_\alpha$, as one sees in the diagrams.  One sees that the
$P=0$ wavefunctions are invariant under rotation by $\pi/2$.  (In fact,
they are angular invariants.)  From the states labeled with nonzero
values of $P$, one can form the complex linear combinations
which transform like $e^{i P \phi}$ when the position and orientation
of the molecule are simultaneously rotated about the symmetry axis.
Although it is far from obvious, the two states labeled
$P=2$ are degenerate in energy.  The two $P=0$ states have different
energy, in general. So quantum mechanics predicts the six state
manifold to consist of two doublets (one for $P= \pm 1$ and one for
$P=\pm 2$) and two singlets (for $P=0$).} 
\label{TRANSROT2}
\end{figure}

\newpage
\begin{figure}
\includegraphics[width=10cm]{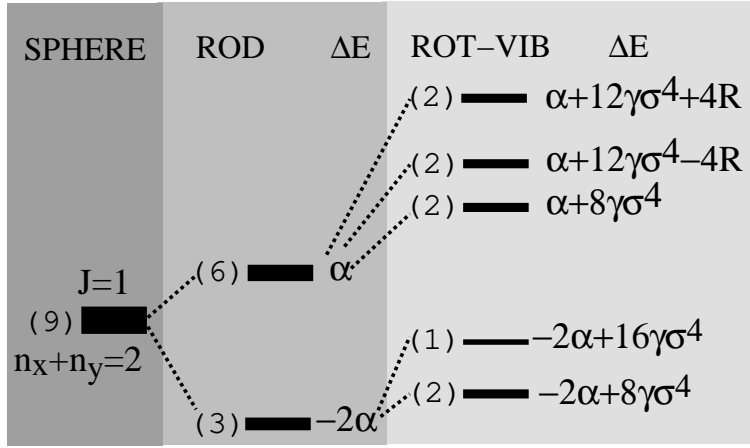}
\vfill
\caption{
Removal of the degeneracy in the energy level
scheme of the two-phonon $(J=1)$ manifold according to the Hamiltonian
of Eq. (\ref{VR1EQ}).  The diagram labeled ``SPHERE" is for a spherical
molecule for which $\alpha=\beta=0$.  That labeled ``ROD" is for
decoupled rotations and translations, but with $\alpha \not=0$.
That labeled ``ROT-VIB" is for anharmonic and  translation-rotation coupling
with $\beta \not=0$.
Here $R = \sqrt{ \gamma^2 \sigma^8 + 2 \beta^2 \sigma^4}$.
}
\label{n2j1levels}
\end{figure}

\newpage
\begin{figure}
\includegraphics[width=10cm]{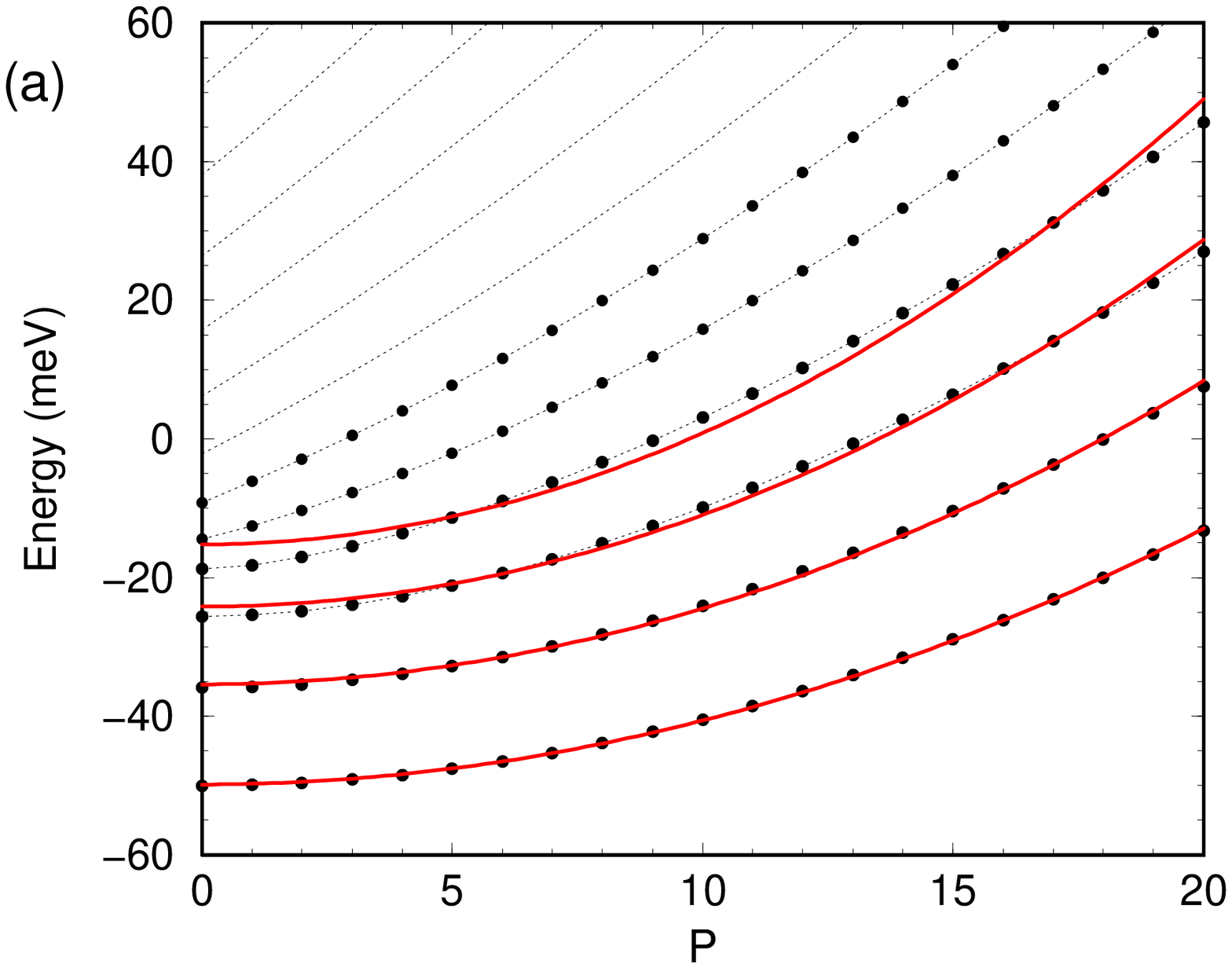}

\includegraphics[width=10cm]{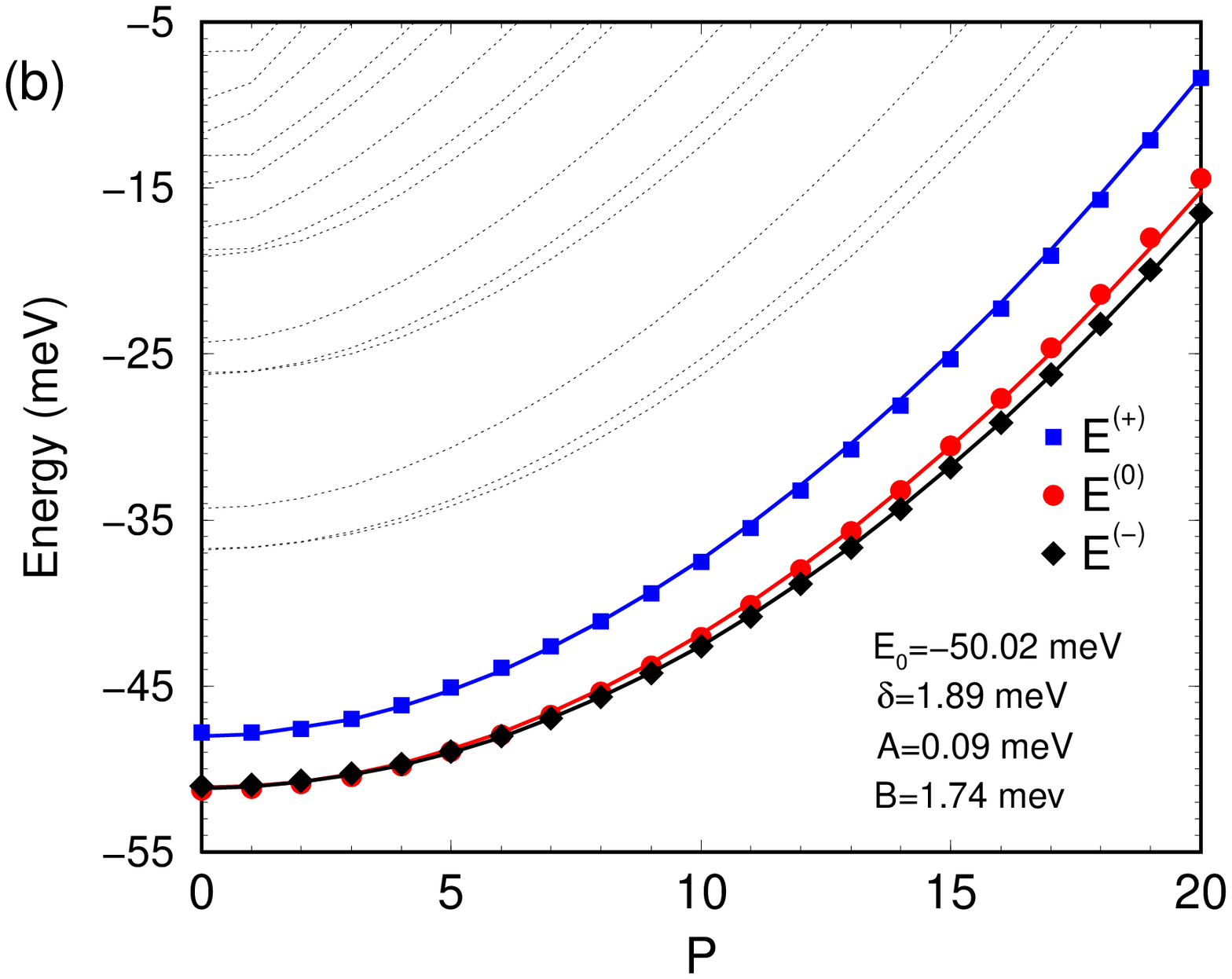}

\vfill
\caption{ The energy levels of of an H$_{2}$ molecule with $(J=0)$ (a)
and $(J=1)$ (b) inside a $(10,10)$ nanotube versus quantum number 
$P$. The symbols and the dotted lines are obtained from 
numerics and the solid lines are fit based on the simple
models as discussed in the text, indicating that 
a few of the lowest energy levels can be understood from these
simple  models.
For the case of $(J=1)$ hydrogen (bottom), for each
$P$ we have now three energies, which are split by about 
3 meV (comparable to 2.6 meV observed for H$_{2}$ on  graphite). 
We note that for both $(J=0)$ and $(J=1)$ cases, only $N=0,1,$ 
and $2$ phonon levels can
be safely identified. The inset to the lower panel gives the
fitted values of the parameters of Eq.~(\ref{SIMPLE}). 
In the lower panel, the energy is with respect to $BJ(J+1)$.
}
\label{10x10}
\end{figure}

\newpage
\begin{figure}
\includegraphics[width=10cm]{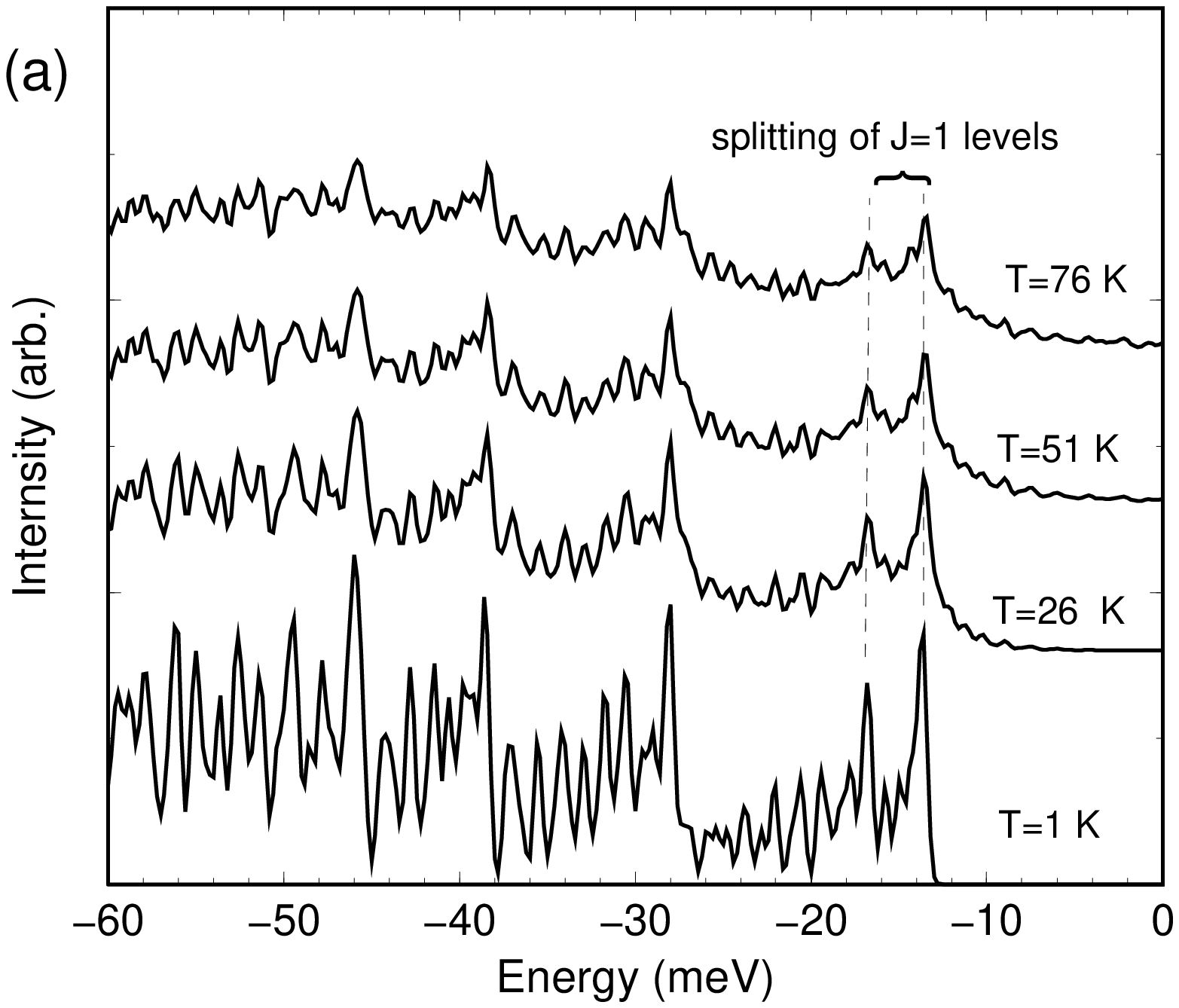}

\includegraphics[width=10cm]{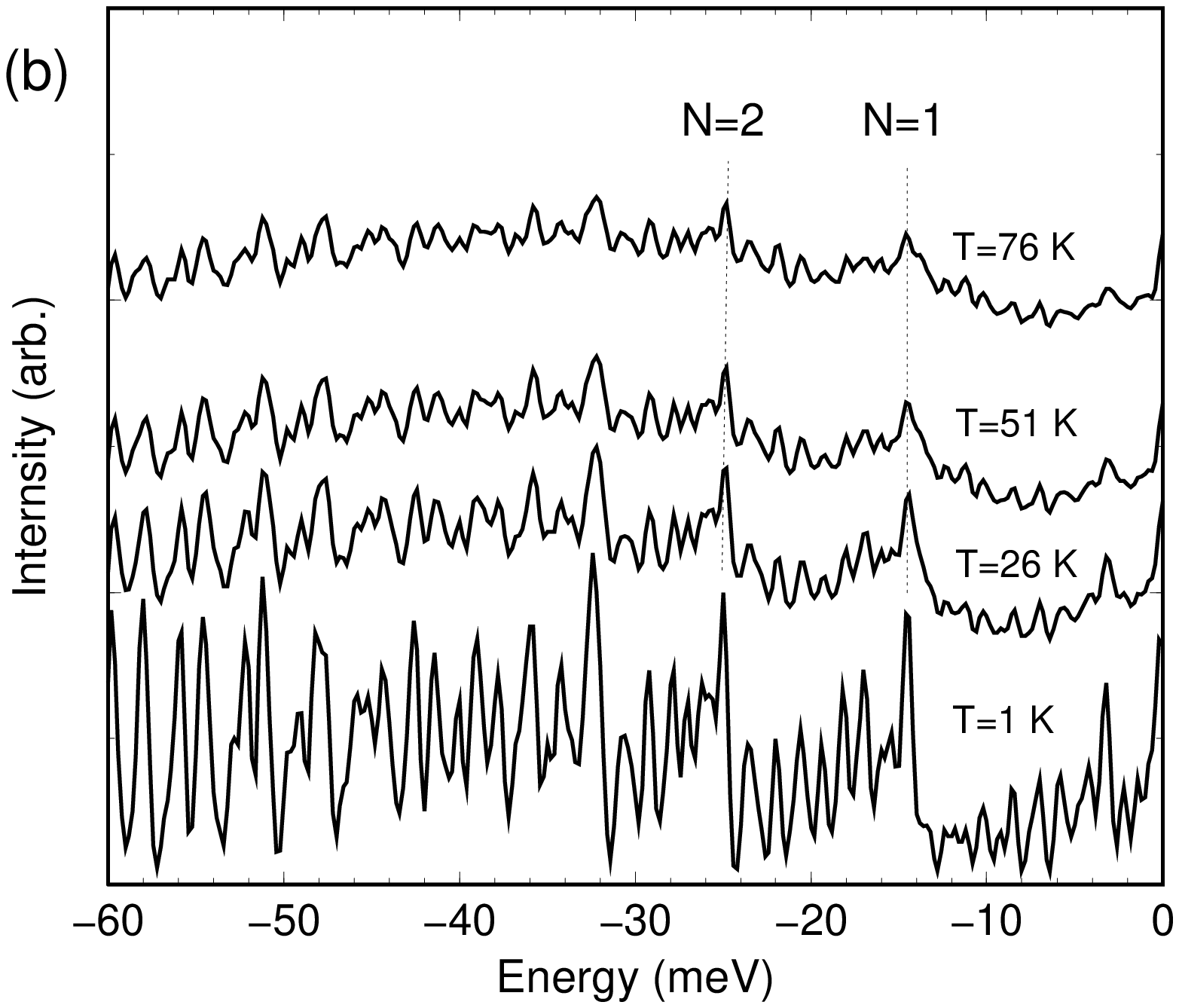}
\vfill 
\caption{The calculated neutron cross section (with arbitrary
scaling)
for para to  ortho (a) and ortho to ortho  (b)  transitions at
several temperatures. The neutron wavevector transfer
${\bf \kappa}$ is taken to be perpendicular to the tube axis
with magnitude of $3$ \AA$^{-1}$.   The peaks are broadened
by Gaussians with FWHM of 0.5 meV. 
}
\label{PHONON}
\end{figure}

\end{document}